\documentclass[aps,prl,preprint,superscriptaddress]{revtex4-2}

\usepackage{graphicx}
\usepackage{amsmath}
\usepackage{bm}
\usepackage{hyperref}
\usepackage{color}
\hypersetup{
	hidelinks,
	colorlinks=true,
	linkcolor=blue,
	urlcolor=blue,
	citecolor=blue,
}

\begin{document}


\title{Experimental Evidence of N\'eel-order-driven Magneto-optical Kerr Effect in an Altermagnetic Insulator}


\author{Haolin Pan}
\affiliation{International Center for Quantum Design of Functional Materials, Hefei National Research Center for Physical Sciences at the Microscale, University of Science and Technology of China, Hefei 230026, China}

\author{Rui-Chun Xiao}
\affiliation{Institute of Physical Science and Information Technology, Anhui University, Hefei 230601, China}
\affiliation{Anhui Provincial Key Laboratory of Magnetic Functional Materials and Devices, School of Materials Science and Engineering, Anhui University, Hefei 230601, China}

\author{Jiahao Han}
\affiliation{Center for Science and Innovation in Spintronics, Tohoku University, Sendai 980-8577, Japan}

\author{Hongxing Zhu}
\author{Junxue Li}
\affiliation{Department of Physics, Southern University of Science and Technology, Shenzhen 518055, China}

\author{Qian Niu}
\affiliation{CAS Key Laboratory of Strongly-Coupled Quantum Matter Physics, University of Science and Technology of China, Hefei, Anhui 230026, China}
\affiliation{Department of Physics, University of Science and Technology of China, Hefei 230026, China}

\author{Yang Gao}
\affiliation{International Center for Quantum Design of Functional Materials, Hefei National Research Center for Physical Sciences at the Microscale, University of Science and Technology of China, Hefei 230026, China}
\affiliation{CAS Key Laboratory of Strongly-Coupled Quantum Matter Physics, University of Science and Technology of China, Hefei, Anhui 230026, China}
\affiliation{Department of Physics, University of Science and Technology of China, Hefei 230026, China}
\affiliation{Anhui Center for Fundamental Sciences in Theoretical Physics, University of Science and Technology of China, Hefei 230026, China}

\author{Dazhi Hou}
	\email[Correspondence author: ]{dazhi@ustc.edu.cn}
\affiliation{International Center for Quantum Design of Functional Materials, Hefei National Research Center for Physical Sciences at the Microscale, University of Science and Technology of China, Hefei 230026, China}
\affiliation{Department of Physics, University of Science and Technology of China, Hefei 230026, China}


\begin{abstract}
The magneto-optical Kerr effect (MOKE) is investigated in hematite, a collinear antiferromagnetic insulator, across a broad wavelength spectrum. By combining the optical measurements with magnetometry results, we unambiguously demonstrate that the N\'eel-order contribution dominates the MOKE signal, while contributions from net magnetization and external magnetic fields are negligible. This conclusion is quantitatively supported by first-principles calculations, and qualitatively by a symmetry analysis that  the N\'eel contribution appears at the first order in spin-orbit coupling while the magnetization contribution  starts only at the third order. This study clarifies the altermagnetic origin of the pronounced MOKE in hematite, underscoring the potential of altermagnets as a promising new class of magneto-optical materials.

\end{abstract}


\maketitle


\textbf{Introduction}---Hematite ($\rm\alpha$-$\rm Fe_2O_3$) is a typical collinear antiferromagnetic (AFM) insulator at room-temperature, exhibiting weak ferromagnetism arising from the Dzyaloshinskii-Moriya (DM) interaction\cite{1960_Moriya_PRL,1960_Moriya_PR}. 
Hematite serves as a model system for exploring intriguing spintronic phenomena in antiferromagnetic insulators \cite{2020_Cheng_PRL_FEO_switching, 2021_Cogulu_PRB_FEO_switching, 2022_Zhang_PRL_FEO_SOT, 2018_Lebrun_Nat_FEO_transport_domain, 2020_Han_FEO_magnon}, owing to its distinctive properties such as high N\'eel temperature \cite{1994_Morrish_Book_FEO}, low damping constant \cite{2018_Lebrun_Nat_FEO_transport_domain, 2020_Han_FEO_magnon}, low magnetocrystalline anisotropy \cite{2020_Cheng_PRL_FEO_switching} and large magneto-optical (MO) response \cite{2021_Cogulu_PRB_FEO_switching}. 
In addition to the conventional second-order MO (Cotton-Mouton) effect typical in collinear AFMs\cite{1969_Pisarev_MLB_1,1969_Pisarev_MLB_2}, hematite displays an unexpectedly large first-order MO signal that is comparable to that in ferromagnets, despite possessing a much weaker net magnetization\cite{1971_Pisarev_MO_0001, 1973_Krinchik_MO_Longitudinal, 1974_Krinchik_MO_Sweep}. 
This phenomenon has been extensively investigated by spectroscopy \cite{1973_Krinchik_MO_Longitudinal, 1981_Zubov_MO_spec, 2020_Ivantsov_MCD} and AFM domain imaging in hematite \cite{1990_Appel_FEO_domain, 2018_Lebrun_Nat_FEO_transport_domain}. 
A widely cited but qualitative explanation attributes the anomalous MOKE in hematite to the N\'eel order under low-symmetry crystal environments, rather than to net magnetization or external magnetic fields—though direct evidence supporting this scenario has remained limited\cite{1994_Morrish_Book_FEO, 2024_Kimel_JMMM_alterm_MO}. 
In early experiments aimed at verifying the N\'eel-order contribution in hematite, the measured MOKE signal contained entangled contributions from surface and volume, making it difficult to reveal the intrinsic response of hematite \cite{1974_Krinchik_MO_Sweep}. More importantly, the contribution of the external magnetic field was completely overlooked without proper verification, resulting in experimental data inadequate to support the proposed explanation \cite{1974_Krinchik_MO_Sweep, 1981_YFeO3}. In the theoretical aspect, it had been challenging to understand how a N\'eel order---with zero net magnetization---can give rise to a MOKE response. Consequently, a definitive conclusion regarding the origin of MOKE in hematite, as well as a microscopic understanding of the underlying mechanism, remains elusive.

The recent discovery of altermagnetism offers a compelling new framework for understanding this anomalous magneto-optical behavior \cite{2022_Smejkal_PRX_alterm}.
Altermagnets, novel collinear AFMs characterized by spin splitting in $k$-space, have become a frontier topic due to their rich physical properties and great potential for spintronic applications \cite{2022_Smejkal_PRX_alterm, 2025_Song_NatRev_altermag}. 
The intrinsic breaking of combined $\hat P \hat T$ symmetry in this material gives rise to ferromagnetic-like effects absent in conventional AFMs, including N\'eel-order-origin anomalous Hall effect (AHE) \cite{2020_Smejkal_SciAdv_alterm_AHE, 2022_Feng_NatElec_RuO2_AHE, 2023_Gonzalez_PRL_MnTe_AHE, 2024_Takagi_NatMat_FeS_AHE}, spin current generation \cite{2021_Gonzalez_PRL_alterm, 2023_Bai_PRL_RuO2_spin} and X-ray magnetic circular dichroism (XMCD) \cite{2023_Lovesey_PRB_MnTe_alterm, 2024_Hariki_PRL_MnTe_XMCD, 2024_Amin_Nat_MnTe_mapping}. Among these, the AHE is usually described by transverse conductivity ($\vec{\sigma}^A$), which is an axial vector representing the antisymmetric part of conductivity tensor and results in the generation of anomalous Hall signal. 
As the optical counterpart of the AHE, a MOKE response originating from the N\'eel order is similarly anticipated in altermagnets \cite{2004_Yao, 2024_Kimel_JMMM_alterm_MO}. 
Notably, hematite is theoretically classified as a $g$-wave altermagnetic insulator \cite{2022_Smejkal_PRX_alterm,2024_Verbeek_PRR_FEO_alterm}, which has been verified in experiments through XMCD and AHE measurements \cite{2025_Galindez_AM}. Recent observation of magnon band splitting in hematite \cite{2023_Kanj_SciAdv_magnon, 2025_Chen_PRL_FEO_magnon} may be also related to its altermagnetism. 
However, direct experimental evidence linking the MOKE in hematite—or in any altermagnetic insulator—to the N\'eel order remains absent.

\begin{figure*}
	\centering
	\includegraphics[width=170mm]{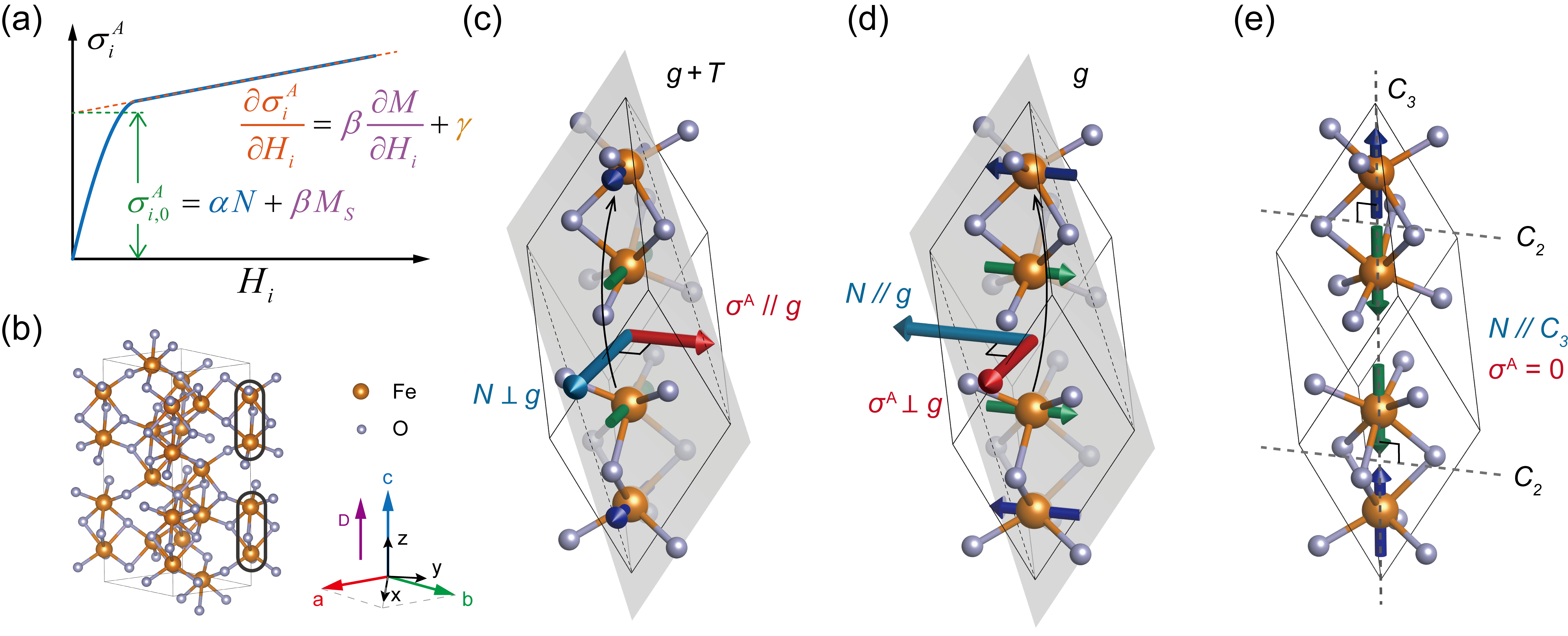}
	\caption{
		(a) Illustration of field-sweeping transverse optical conductivity $\sigma^A_{i}$ results. 
		(b) Crystal structure of hematite in hexagonal lattice and definition of Cartesian coordinate system $xyz$, where $x\parallel [11\bar{2}0]$, $y\parallel [\bar{1}100]$ and $z\parallel [0001]$.
		(c), (d), (e) Magnetic symmetries and corresponding $\vec\sigma^A$ of hematite induced by N\'eel order in rhombohedral lattice, when the $\vec N$ is oriented along $x$, $y$ and $z$ directions. The relative positions of Fe atoms are marked in (b). The glide plane $g$ consists of mirror reflection against the gray plane and translation along the $c$ axis, denoted by black arrows. $\hat T$, $C_3$ and $C_2$ represent the time reversal, three-fold rotation and two-fold rotation, respectively.
	}
	\label{fig1}  
\end{figure*}

The primary challenge in elucidating the origin of the MOKE in hematite arises because the transverse optical conductivity $\vec{\sigma}^{A}$ could have contributions from three distinct sources \cite{2024_Kimel_JMMM_alterm_MO,2024_Xiao_altermag}: N\'eel vector $\vec{N}$, net magnetization $\vec{M}$, and external magnetic field $\vec{H}$, given
\begin{equation}
	\vec{\sigma}^{A}=\boldsymbol{\alpha}\cdot\vec{N} +\boldsymbol{\beta}\cdot\vec{M}+ \boldsymbol{\gamma}\cdot\vec{H}, 
	\label{eq:init}
\end{equation}
where $\boldsymbol{\alpha}$, $\boldsymbol{\beta}$ and $\boldsymbol{\gamma}$ are rank-2 tensors, while the MOKE signal offers only two measurable quantities—the zero-field intercept $\sigma^A_{i,0}$ and the field slope $\partial\sigma^A_{i}/\partial H_i$—as illustrated in FIG.~\ref{fig1}a, making it difficult to disentangle the underlying contributions \cite{2024_Kimel_JMMM_alterm_MO}. 
In Eq.~\ref{eq:init}, only first order contributions are considered. 
Specifically, in collinear AFMs, $\vec{M}$ includes two distinct components: spontaneous magnetization ($M_S$) and field-induced magnetization ($M_H$). The MO contribution from $M_S$ remains constant in a single-domain state and is thus inseparable from the N\'eel-vector contribution within the intercept $\sigma^A_{i,0}$. Conversely, the MO contribution from $M_H$ is linearly proportional to $H$ provided the applied field remains significantly below the exchange field (approximately 970 T for hematite \cite{1994_Morrish_Book_FEO}), thus blending with the $\vec{H}$ contribution in the slope $\partial\sigma^A_i/\partial H_i$. 
Consequently, with three unknowns in only two equations, we cannot confirm the existence of the N\'eel-order contribution.  
This issue is quite common in collinear antiferromagnets \cite{1974_Krinchik_MO_Sweep, 1981_YFeO3}. 
To address this issue, previous AHE studies on FeS assumed negligible magnetization contribution based on \textit{ab-initio} calculations, enabling clear separation of N\'eel and external-field contributions across temperature ranges\cite{2024_Takagi_NatMat_FeS_AHE}. 
Therefore, the idea that the MOKE signal in hematite originates from the N\'eel order has remained a hypothesis. Whether this signal contains a significant contribution from net magnetization is also unresolved. 
To address this difficulty, we take advantage of the photon energy---a unique parameter in optical measurement---to resolve these problem. By systematically analyzing the photon-energy-dependent behavior of both intercept and slope across hematite samples with varied orientations, we rigorously identify the N\'eel order as the dominant contributor to the MOKE response, which is a hallmark feature of altermagnetism.

\textbf{Symmetry Analysis}---Before presenting the results of the MOKE measurement, we perform a brief symmetry analysis on hematite with three typical N\'eel order orientations to clarify $\vec{\sigma}^{A}$ originated from altermagnetism. 
The Cartesian axes $xyz$ used here is defined with respect to the hexagonal lattice of hematite, as shown in FIG.~\ref{fig1}b. 
Hematite has space group of $R\bar 3c$ and spin arrangement of $+--+$ in rhombohedral lattice \cite{1994_Morrish_Book_FEO,2008_Hill_FEO_ND}. 
For simplification, we consider spin orderings with zero-canting in hematite. 
For $\vec{N}\parallel x$ (FIG.~\ref{fig1}c), the glide plane $g$ preserves the crystal structure but reverses $\vec{N}$. 
Consequently, the system exhibits symmetry under time reversal $\hat T$ followed by $g$, which constrains $\vec{\sigma}^A$ parallel to $g$, since $\vec\sigma^A\parallel g$ keeps invariant under the $g\hat T$ operation while $\vec\sigma^A\perp g$ reverses. 
For $\vec{N}\parallel y$ (FIG.~\ref{fig1}d), $g$ preserves both the crystal structure and $\vec{N}$, therefore confining $\vec{\sigma}^A$ perpendicular to $g$. 
When $\vec{N}$ orients along the $c$-axis (FIG.~\ref{fig1}e), the simultaneous preservation of $C_3$ and $C_2$ rotational symmetry forces $\vec{\sigma}^A$ to be parallel to both rotation axes and therefore vanishing. 
Taking all the cases into consideration, $\boldsymbol{\alpha}\cdot\vec{N}$ can be simplified as $\alpha\cdot(\vec{c}\times\vec{N})$ \cite{2024_Xiao_altermag}, where $\alpha$ is a scalar and $\vec{c}$ is the unit vector in $c$-axis.
Although $\vec{\sigma}^A$ induced by $\vec{M}$ and $\vec{H}$ have no symmetry requirements, the spontaneous magnetization itself is relevant to crystal symmetry through DM interaction. 
As established\cite{1994_Morrish_Book_FEO,2024_Kimel_JMMM_alterm_MO,2025_Roig_PRL}, $\vec{M}_S$ can be expressed as $\vec{D}\times \vec{N} = D\cdot(\vec{c}\times \vec{N})$, where vector $\vec{D}$ parallel to the $c$-axis is the first order approximation of DM interaction. 
$\vec{M}_H$ can be expressed as $\mu\vec{H}$, since the magnetic susceptibilities in the $ab$-plane and along $c$-axis are nearly equal as reported \cite{1994_Morrish_Book_FEO, 2023_Kanj_SciAdv_magnon}.
Therefore, the total expression of transverse optical conductivity becomes: 
\begin{equation}
	\vec{\sigma}^{A} = \boldsymbol{\alpha}\cdot\vec{N} +\boldsymbol{\beta}\cdot(D\cdot\vec{c}\times\vec{N}+\mu\vec{H})+ \boldsymbol{\gamma}\cdot\vec{H} =(\alpha+\boldsymbol{\beta}D) \cdot (\vec{c}\times\vec{N}) +(\boldsymbol{\beta}\mu +\boldsymbol{\gamma})\cdot \vec{H},
	\label{eq:general}
\end{equation}
demonstrating the mixing effects of $\vec N$ and $\vec M_S$ for arbitrary $\vec N$ direction. Therefore, the contributions of $N$ and $M_S$ can not be distinguished through symmetry consideration; quantitative analysis is necessary to solve this problem. 

\begin{figure}
	\centering
	\includegraphics[width=85mm]{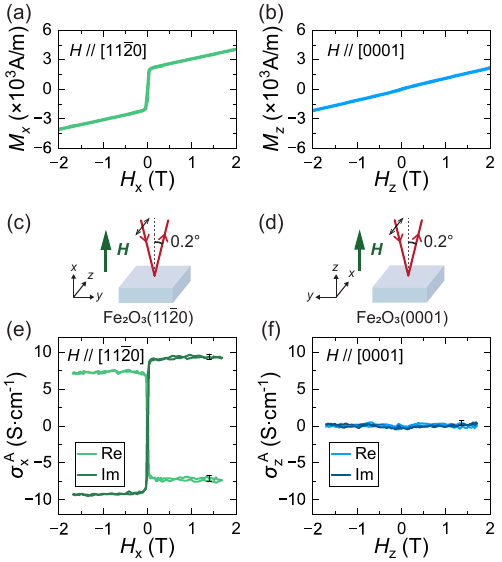}
	\caption{
		(a), (b) Magnetic hysteresis of hematite along the $x$ and $z$ directions at room temperature. 
		(c), (d) MOKE configuration of hematite $(11\bar{2}0)$ and $(0001)$ samples. The incident angle is 0.2 deg. 
		(e), (f) $\sigma^A_{x}$ and $\sigma^A_{z}$ of hematite at the wavelength of 550 nm, when $H$ is applied along $x$ and $z$ directions, respectively. The length of error bar is 0.72 S/cm.
	}
	\label{fig2}  
\end{figure}

\textbf{Sample Characterization}---The experiments are conducted in hematite $(11\bar{2}0)$ and $(0001)$ bulk samples. 
X-ray diffraction characterizations confirm high monocrystallinity in both samples \cite{supp}.
The magnetic properties of samples in out-of-plane direction, corresponding to the $x$ and $z$ directions of hematite, are measured by vibrating sample magnetometer at room temperature, as shown in FIG.~\ref{fig2}a and \ref{fig2}b. 
The $(11\bar{2}0)$ sample enters single domain state under magnetic field of 0.2 T along $x$-axis, indicating $ab$-plane as easy plane for $\vec N$ with low magnetocrystalline anisotropy. 
The $M_S$ of the sample in the $x$ direction is $2.13\times10^3$ A/m. By adopting sublattice magnetization as $7.75\times10^5$ A/m \cite{2008_Hill_FEO_ND}, the DM coefficient $D$ is determined as $1.37\times 10^{-3}$. 
The $(0001)$ sample shows near zero spontaneous magnetization in the $z$ direction. 
In this case, $\vec N$ lies along equivalent $[1\bar{1}00]$ directions, which correspond to the relatively easy axes within the $ab$-plane.
The magnetic susceptibilities are $1.22\times10^{-3}$ along $x$-axis and $1.31\times10^{-3}$ along $z$-axis, respectively. 
These results demonstrate good agreement with previous report \cite{1994_Morrish_Book_FEO}.

\textbf{H-sweeping MOKE}---The polar MOKE signal with out-of-plane magnetic field is measured for each sample, as illustrated in FIG.~\ref{fig2}c and \ref{fig2}d. 
The procedure for the MOKE measurement follows previous works \cite{1985_Querry_N,1972_Johnson_N,1998_You_formula,2000_Qiu_MOreview}, and the data analysis is included in the Supplemental Material \cite{supp}.
The incident angle of linearly polarized light to samples is set as 0.2 deg, which is a good approximation of normal incidence. 
The rotation and ellipticity of light are measured using the cross-polarizer method, and then converted to transverse optical conductivity $\sigma^A_{x}$ and $\sigma^A_{z}$ in $(11\bar{2}0)$ and $(0001)$ sample, respectively.
According to Eq.~\ref{eq:general}, the dependence of $\sigma^A_{x}$ on $H_x$ can be expressed as follows: 
\begin{equation}
	\sigma^A_{x} = (\alpha+\beta_{xx}D)N_y +(\beta_{xx}\mu+\gamma_{xx})H_x,
	\label{eq:x}
\end{equation}
while the dependence of $\sigma^A_{z}$ on $H_z$ has expression: 
\begin{equation}
	\sigma^A_{z} = (\beta_{zz}\mu+\gamma_{zz})H_z.
	\label{eq:z}
\end{equation}

The field-sweeping results of $\sigma^A_{x}$ and $\sigma^A_{z}$ at the wavelength of 550 nm are shown in FIG.~\ref{fig2}e and \ref{fig2}f, both of which exhibit near-zero slope above 0.2 T. 
The intercept of $\sigma^A_{x}$ in the single-domain region is $(-7.31 + 9.21i)\ \rm{S/cm}$ while that of $\sigma^A_{z}$ is $( 0.06 + 0.08i)\ \rm{S/cm}$, corresponding to the description of Eq.~\ref{eq:x} and Eq.~\ref{eq:z}. 
The large intercept corresponds to the magnitude of 1.7 mrad in MOKE signal, which is comparable to typical ferromagnets Fe and Co \cite{2018_Higo_MOKE}.
The slopes of $\sigma^A_{x}$ and $\sigma^A_{z}$ are $(4.15 +5.88i)\times 10^{-2}\ \rm{S/(cm\cdot T)}$ and $(-4.34 +6.18i)\times 10^{-2}\ \rm{S/(cm\cdot T)}$, which are below the noise level of the measurements. 

Further analysis can be conducted by assuming a negligible contribution from the external field, specifically that $\gamma_{xx}\approx\gamma_{zz}\approx 0$. Under this assumption, the slope of the MOKE signal arises exclusively from the net magnetization. Consequently, the coefficients $\beta_{xx}$ and $\beta_{zz}$ can be determined as $(4.28 + 6.06i)\times 10^{-3}\ \rm{S/A}$ and $(-4.17 + 5.94i)\times 10^{-3}\ \rm{S/A}$, respectively. By subtracting the magnetization contribution from the intercept based on the data in FIG.~\ref{fig1}a, the coefficient $\alpha$ is found to be $(-4.77 + 5.86i)\times 10^{-4}\ \rm{S/A}$. Given that $|\alpha N|$ is two orders of magnitude larger than $|\beta_{xx} M_S|$, it clearly indicates that the MOKE signal in hematite predominantly originates from the $N$, with negligible contributions from $M_S$.

Alternatively, without assuming a negligible external-field contribution, the observed near-zero slope could result from a coincidental cancellation between finite contributions from $M_H$ and $H$. Considering $M_H$ and $M_S$ are comparable in our measurements, the MO contribution of $M_S$ can also be finite, leading to ambiguity in distinguishing the proportions contributed by the $N$ and $M_S$. However, since the MO contributions from $M_H$ and $H$ originate from fundamentally different physical mechanisms and should exhibit distinct photon-energy dependencies, such cancellations would be purely accidental and unlikely to persist consistently across the entire spectral range.

\begin{figure*}
	\centering
	\includegraphics[width=170mm]{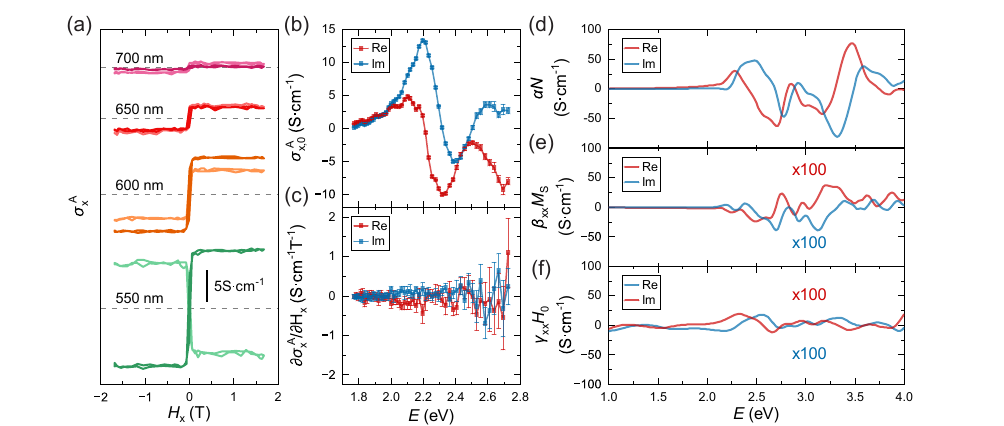}
	\caption{
		(a) $H_x$ dependency of $\sigma^A_{x}$ at different wavelengths, measured in the hematite $(11\bar{2}0)$ sample. Light lines correspond to the real part, while dark lines correspond to the imaginary part.
		(b),(c) Intercept and slope of $\sigma^A_{x}$ spectrum. 
		(d),(e),(f) Calculated N\'eel-order-induced conductivity $\alpha N$, magnetization-induced conductivity $\beta_{xx}M_S$ and field-induced conductivity $\gamma_{xx}H_0$, where $M_S=2.13\times10^3\ \rm A/m$, $H_0=1\ \rm T$.
	}
	\label{fig3}  
\end{figure*}

\textbf{Spectrum Analysis}---The field-sweeping results of $\sigma^A_{x}$ at different wavelengths, measured in hematite $(11\bar{2}0)$ sample, are shown in FIG.~\ref{fig3}a. 
All the curves exhibit a consistent zero slope in the single-domain region. 
The intercept ($\sigma^A_{x,0}$) and slope ($\partial\sigma^A_{x} /\partial H_x$) in the visible light region are shown in FIG.~\ref{fig3}b and \ref{fig3}c, with the standard errors of the linear fits represented as error bars. 
According to Eq.~\ref{eq:x}, the magnitude of $\partial\sigma^A_{x} /\partial H_x$, i.e., $|\beta_{xx}\mu +\gamma_{xx}|$, remains constantly smaller than 0.5 $\rm{S/(cm\cdot T)}$, except for points above 2.6 eV due to the reduced intensity of the light source. 
Since $\beta_{xx}$ and $\gamma_{xx}$ have distinct dependencies on photon energy, the consistently near zero slope of $\sigma^A_{x}$ indicates that both $\beta_{xx}$ and $\gamma_{xx}$ are too small to be observed. 
Consequently, we estimate that the upper limits of $|\beta_{xx}\mu|$ and $|\gamma_{xx}|$ are 0.5 $\rm{S/(cm\cdot T)}$ within the visible light range. 
This indicates that the contribution of $M_S$ to $\sigma^A_{x,0}$ should be less than 1 $\rm{S/cm}$, which amounts to a small fraction of measured values in FIG.~\ref{fig3}b. 

\begin{figure}
	\centering
	\includegraphics[width=85mm]{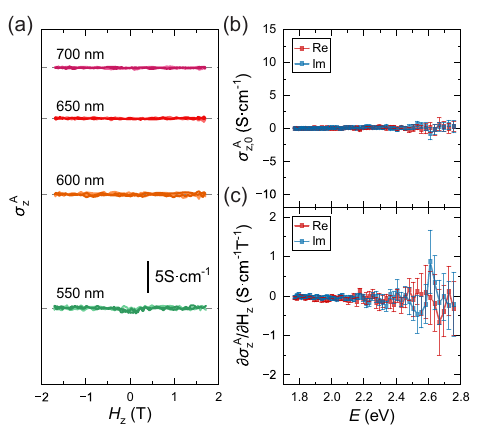}
	\caption{
		(a) $H_z$ dependency of $\sigma^A_{z}$ at different wavelengths, measured in the hematite $(0001)$ sample. Light lines correspond to the real part, while dark lines correspond to the imaginary part. 
		(b),(c) Intercept and slope of $\sigma^A_{z}$ spectrum. 
	}
	\label{fig4}  
\end{figure}

Our conclusion can be further validated by the MOKE measurements in the hematite $(0001)$ sample. 
The field-sweeping results for $\sigma^A_{z}$ at different wavelengths, shown in FIG.~\ref{fig4}a, also exhibit zero slopes in the single-domain region.
The values of $\sigma^A_{z,0}$ shown in FIG.~\ref{fig4}b are close to zero in the visible light region, consistent with Eq.~\ref{eq:z}. 
The magnitude of $\partial\sigma^A_{z} /\partial H_z$ is consistently smaller than 0.3 $\rm{S/(cm\cdot T)}$, except for a few noisy data points at higher energies. 
These results suggest that the magnitudes of $\beta_{zz}\mu$ and $\gamma_{zz}$ are smaller than 0.3 $\rm{S/(cm\cdot T)}$, negligible compared with MOKE induced by N\'eel order. 

\textbf{Theoretical Calculation}---The dominance of the N\'eel-order contribution in the MOKE of hematite is further corroborated by our first-principles calculations on each contributions. 
The details of first-principles calculations and formula derivations are included in the Supplemental Material \cite{supp} (see also references  \cite{1996_Ebert_RPP,2004_Yao,2016_Xiao_PRL,2008_Mostofi_CPC,1996_Kresse_PRB, 1999_Kresse_PRB,2010_Book_Condensed,2026_Xiao} therein). 
The anomalous mechanism of the MOKE is from the transverse optical conductivity. The general optical conductivity is defined as
\begin{align}
	J_i=\sigma_{ij}(\omega)E_j\,,
\end{align}
where $\sigma_{ij}(\omega)$ can be obtained using the Kubo-Greenwood formula \cite{1996_Ebert_RPP,2004_Yao,2016_Xiao_PRL}. The transverse optical conductivity is then the antisymmetric part of $\sigma_{ij}(\omega)$, i.e., $\sigma^A_i=\frac{1}{2}\epsilon_{ijk}\sigma_{jk}$, where $\epsilon_{ijk}$ is the Levi-Civita symbol. 
Like the anomalous Hall effect, the transverse optical conductivity can be viewed as a spin-group symmetry breaking phenomenon \cite{2025_Liu_PRX}, and the N\'eel-order contribution occurs at linear order of the spin-orbit coupling (SOC) strength. In contrast, the $\vec M_S$ itself scales linearly with DM interaction, which is of the second order in the SOC \cite{1960_Moriya_PRL,1960_Moriya_PR}, so its contribution to MOKE signal shall start at the third order.
By constraining the direction of the local spin order in the first-principles calculations, we can calculate $\sigma_{i}^A(\omega)$ originating from the N\'eel order, $\alpha N$, and spontaneous magnetization, $\beta_{xx}M_S$, as shown in FIG.~\ref{fig3}d and \ref{fig3}e. 
Clearly, the N\'eel-order contribution is much larger. The trend of $\alpha N$ matches well with the experimental values of $\sigma^A_{x,0}$. To further evaluate the magnetic-field induced MOKE, we consider the following second-order response function
\begin{align}
	J_i=\sigma_{ijk}(\omega)E_jB_k\,,
\end{align}
and using the density-matrix perturbation theory, we can derive the expression of $\sigma_{ijk}(\omega)$. We then take the antisymmetrization of the $ij$ indices and evaluate the resulting expression using first-principles calculations. The result is shown in FIG.~\ref{fig3}f with $H_0=1\ \rm T$, which is also much smaller than the N\'eel-order contribution. This can be understood that magnetic field effect is much smaller than the exchange coupling associated with the N\'eel order.

\textbf{Summary}---We clarify that the dominant contribution to the large MOKE signal in hematite is from the N\'eel order, which is a hallmark of the altermagnetic behavior in this material. The contributions from both the net magnetization and the external magnetic field are found to be negligible within the visible light range, which is verified in experiment and first-principles calculation for the first time. Our investigation demonstrates the ability of altermagnets to generate large MO signal, opening rich opportunities in the emergent field of altermagnetic opto-spintronics,  including the exploration of altermagnetic insulator candidates like $\rm YFeO_3$, $\rm LaMnO_3$ and $\rm LaCrO_3$ \cite{2018_Higo_MOKE,2025_Naka_perovskites} and the development of altermagnet-based optical devices like zero-field optical isolators and integrated photonic logics gates \cite{2022_rev_isolator,2025_Lapointe_isolator,2025_Pintus_logic}.

\textit{Note added}: During the review process, we are aware that the work of J. Lou et al. \cite{2025_Luo_arxiv} reaches similar conclusions.

\section{ACKNOWLEDGMENTS}
\begin{acknowledgments}
	This work was supported by the National Key R\&D Program under grant Nos. 2022YFA1403502, the National Natural Science Foundation of China (12234017, 12074366, 12374164). D. Hou and Y. Gao were supported by the Fundamental Research Funds for the Central Universities (Grant No. WK9990000116, WK2340000102). R. Xiao was supported by the National Natural Science Foundation of China (Grants Nos. 12474100, 12204009). H. Zhu and J. Li were supported by the National Natural Science Foundation of China (No. 12374111). This work was supported by the USTC Center for Micro- and Nanoscale Research and Fabrication. This work was partially carried out at Instruments Center for Physical Science, University of Science and Technology of China. 
\end{acknowledgments}



%


\begin{thebibliography}{56}%
	\makeatletter
	\providecommand \@ifxundefined [1]{%
		\@ifx{#1\undefined}
	}%
	\providecommand \@ifnum [1]{%
		\ifnum #1\expandafter \@firstoftwo
		\else \expandafter \@secondoftwo
		\fi
	}%
	\providecommand \@ifx [1]{%
		\ifx #1\expandafter \@firstoftwo
		\else \expandafter \@secondoftwo
		\fi
	}%
	\providecommand \natexlab [1]{#1}%
	\providecommand \enquote  [1]{``#1''}%
	\providecommand \bibnamefont  [1]{#1}%
	\providecommand \bibfnamefont [1]{#1}%
	\providecommand \citenamefont [1]{#1}%
	\providecommand \href@noop [0]{\@secondoftwo}%
	\providecommand \href [0]{\begingroup \@sanitize@url \@href}%
	\providecommand \@href[1]{\@@startlink{#1}\@@href}%
	\providecommand \@@href[1]{\endgroup#1\@@endlink}%
	\providecommand \@sanitize@url [0]{\catcode `\\12\catcode `\$12\catcode
		`\&12\catcode `\#12\catcode `\^12\catcode `\_12\catcode `\%12\relax}%
	\providecommand \@@startlink[1]{}%
	\providecommand \@@endlink[0]{}%
	\providecommand \url  [0]{\begingroup\@sanitize@url \@url }%
	\providecommand \@url [1]{\endgroup\@href {#1}{\urlprefix }}%
	\providecommand \urlprefix  [0]{URL }%
	\providecommand \Eprint [0]{\href }%
	\providecommand \doibase [0]{https://doi.org/}%
	\providecommand \selectlanguage [0]{\@gobble}%
	\providecommand \bibinfo  [0]{\@secondoftwo}%
	\providecommand \bibfield  [0]{\@secondoftwo}%
	\providecommand \translation [1]{[#1]}%
	\providecommand \BibitemOpen [0]{}%
	\providecommand \bibitemStop [0]{}%
	\providecommand \bibitemNoStop [0]{.\EOS\space}%
	\providecommand \EOS [0]{\spacefactor3000\relax}%
	\providecommand \BibitemShut  [1]{\csname bibitem#1\endcsname}%
	\let\auto@bib@innerbib\@empty
	\bibitem [{\citenamefont {Moriya}(1960{\natexlab{a}})}]{1960_Moriya_PRL}%
	\BibitemOpen
	\bibfield  {author} {\bibinfo {author} {\bibfnamefont {T.}~\bibnamefont
			{Moriya}},\ }\bibfield  {title} {\bibinfo {title} {New mechanism of
			anisotropic superexchange interaction},\ }\href
	{https://doi.org/10.1103/PhysRevLett.4.228} {\bibfield  {journal} {\bibinfo
			{journal} {Phys. Rev. Lett.}\ }\textbf {\bibinfo {volume} {4}},\ \bibinfo
		{pages} {228} (\bibinfo {year} {1960}{\natexlab{a}})}\BibitemShut {NoStop}%
	\bibitem [{\citenamefont {Moriya}(1960{\natexlab{b}})}]{1960_Moriya_PR}%
	\BibitemOpen
	\bibfield  {author} {\bibinfo {author} {\bibfnamefont {T.}~\bibnamefont
			{Moriya}},\ }\bibfield  {title} {\bibinfo {title} {Anisotropic superexchange
			interaction and weak ferromagnetism},\ }\href
	{https://doi.org/10.1103/PhysRev.120.91} {\bibfield  {journal} {\bibinfo
			{journal} {Phys. Rev.}\ }\textbf {\bibinfo {volume} {120}},\ \bibinfo {pages}
		{91} (\bibinfo {year} {1960}{\natexlab{b}})}\BibitemShut {NoStop}%
	\bibitem [{\citenamefont {Cheng}\ \emph {et~al.}(2020)\citenamefont {Cheng},
		\citenamefont {Yu}, \citenamefont {Zhu}, \citenamefont {Hwang},\ and\
		\citenamefont {Yang}}]{2020_Cheng_PRL_FEO_switching}%
	\BibitemOpen
	\bibfield  {author} {\bibinfo {author} {\bibfnamefont {Y.}~\bibnamefont
			{Cheng}}, \bibinfo {author} {\bibfnamefont {S.}~\bibnamefont {Yu}}, \bibinfo
		{author} {\bibfnamefont {M.}~\bibnamefont {Zhu}}, \bibinfo {author}
		{\bibfnamefont {J.}~\bibnamefont {Hwang}},\ and\ \bibinfo {author}
		{\bibfnamefont {F.}~\bibnamefont {Yang}},\ }\bibfield  {title} {\bibinfo
		{title} {Electrical switching of tristate antiferromagnetic {N\'eel} order in
			$\ensuremath{\alpha}$\text{\ensuremath{-}}{$\rm Fe_2O_3$} epitaxial films},\
	}\href {https://doi.org/10.1103/PhysRevLett.124.027202} {\bibfield  {journal}
		{\bibinfo  {journal} {Phys. Rev. Lett.}\ }\textbf {\bibinfo {volume} {124}},\
		\bibinfo {pages} {027202} (\bibinfo {year} {2020})}\BibitemShut {NoStop}%
	\bibitem [{\citenamefont {Cogulu}\ \emph {et~al.}(2021)\citenamefont {Cogulu},
		\citenamefont {Statuto}, \citenamefont {Cheng}, \citenamefont {Yang},
		\citenamefont {Chopdekar}, \citenamefont {Ohldag},\ and\ \citenamefont
		{Kent}}]{2021_Cogulu_PRB_FEO_switching}%
	\BibitemOpen
	\bibfield  {author} {\bibinfo {author} {\bibfnamefont {E.}~\bibnamefont
			{Cogulu}}, \bibinfo {author} {\bibfnamefont {N.~N.}\ \bibnamefont {Statuto}},
		\bibinfo {author} {\bibfnamefont {Y.}~\bibnamefont {Cheng}}, \bibinfo
		{author} {\bibfnamefont {F.}~\bibnamefont {Yang}}, \bibinfo {author}
		{\bibfnamefont {R.~V.}\ \bibnamefont {Chopdekar}}, \bibinfo {author}
		{\bibfnamefont {H.}~\bibnamefont {Ohldag}},\ and\ \bibinfo {author}
		{\bibfnamefont {A.~D.}\ \bibnamefont {Kent}},\ }\bibfield  {title} {\bibinfo
		{title} {Direct imaging of electrical switching of antiferromagnetic {N\'eel}
			order in
			$\ensuremath{\alpha}\text{\ensuremath{-}}{{\mathrm{Fe}}_{2}\mathrm{O}}_{3}$
			epitaxial films},\ }\href {https://doi.org/10.1103/PhysRevB.103.L100405}
	{\bibfield  {journal} {\bibinfo  {journal} {Phys. Rev. B}\ }\textbf {\bibinfo
			{volume} {103}},\ \bibinfo {pages} {L100405} (\bibinfo {year}
		{2021})}\BibitemShut {NoStop}%
	\bibitem [{\citenamefont {Zhang}\ \emph {et~al.}(2022)\citenamefont {Zhang},
		\citenamefont {Chou}, \citenamefont {Yun}, \citenamefont {McGoldrick},
		\citenamefont {Hou}, \citenamefont {Mkhoyan},\ and\ \citenamefont
		{Liu}}]{2022_Zhang_PRL_FEO_SOT}%
	\BibitemOpen
	\bibfield  {author} {\bibinfo {author} {\bibfnamefont {P.}~\bibnamefont
			{Zhang}}, \bibinfo {author} {\bibfnamefont {C.-T.}\ \bibnamefont {Chou}},
		\bibinfo {author} {\bibfnamefont {H.}~\bibnamefont {Yun}}, \bibinfo {author}
		{\bibfnamefont {B.~C.}\ \bibnamefont {McGoldrick}}, \bibinfo {author}
		{\bibfnamefont {J.~T.}\ \bibnamefont {Hou}}, \bibinfo {author} {\bibfnamefont
			{K.~A.}\ \bibnamefont {Mkhoyan}},\ and\ \bibinfo {author} {\bibfnamefont
			{L.}~\bibnamefont {Liu}},\ }\bibfield  {title} {\bibinfo {title} {Control of
			{N\'eel} vector with spin-orbit torques in an antiferromagnetic insulator
			with tilted easy plane},\ }\href
	{https://doi.org/10.1103/PhysRevLett.129.017203} {\bibfield  {journal}
		{\bibinfo  {journal} {Phys. Rev. Lett.}\ }\textbf {\bibinfo {volume} {129}},\
		\bibinfo {pages} {017203} (\bibinfo {year} {2022})}\BibitemShut {NoStop}%
	\bibitem [{\citenamefont {Lebrun}\ \emph {et~al.}(2018)\citenamefont {Lebrun},
		\citenamefont {Ross}, \citenamefont {Bender}, \citenamefont {Qaiumzadeh},
		\citenamefont {Baldrati}, \citenamefont {Cramer}, \citenamefont {Brataas},
		\citenamefont {Duine},\ and\ \citenamefont
		{Kl{\"a}ui}}]{2018_Lebrun_Nat_FEO_transport_domain}%
	\BibitemOpen
	\bibfield  {author} {\bibinfo {author} {\bibfnamefont {R.}~\bibnamefont
			{Lebrun}}, \bibinfo {author} {\bibfnamefont {A.}~\bibnamefont {Ross}},
		\bibinfo {author} {\bibfnamefont {S.~A.}\ \bibnamefont {Bender}}, \bibinfo
		{author} {\bibfnamefont {A.}~\bibnamefont {Qaiumzadeh}}, \bibinfo {author}
		{\bibfnamefont {L.}~\bibnamefont {Baldrati}}, \bibinfo {author}
		{\bibfnamefont {J.}~\bibnamefont {Cramer}}, \bibinfo {author} {\bibfnamefont
			{A.}~\bibnamefont {Brataas}}, \bibinfo {author} {\bibfnamefont {R.~A.}\
			\bibnamefont {Duine}},\ and\ \bibinfo {author} {\bibfnamefont
			{M.}~\bibnamefont {Kl{\"a}ui}},\ }\bibfield  {title} {\bibinfo {title}
		{Tunable long-distance spin transport in a crystalline antiferromagnetic iron
			oxide},\ }\href {https://doi.org/10.1038/s41586-018-0490-7} {\bibfield
		{journal} {\bibinfo  {journal} {Nature}\ }\textbf {\bibinfo {volume} {561}},\
		\bibinfo {pages} {222} (\bibinfo {year} {2018})}\BibitemShut {NoStop}%
	\bibitem [{\citenamefont {Han}\ \emph {et~al.}(2020)\citenamefont {Han},
		\citenamefont {Zhang}, \citenamefont {Bi}, \citenamefont {Fan}, \citenamefont
		{Safi}, \citenamefont {Xiang}, \citenamefont {Finley}, \citenamefont {Fu},
		\citenamefont {Cheng},\ and\ \citenamefont {Liu}}]{2020_Han_FEO_magnon}%
	\BibitemOpen
	\bibfield  {author} {\bibinfo {author} {\bibfnamefont {J.}~\bibnamefont
			{Han}}, \bibinfo {author} {\bibfnamefont {P.}~\bibnamefont {Zhang}}, \bibinfo
		{author} {\bibfnamefont {Z.}~\bibnamefont {Bi}}, \bibinfo {author}
		{\bibfnamefont {Y.}~\bibnamefont {Fan}}, \bibinfo {author} {\bibfnamefont
			{T.~S.}\ \bibnamefont {Safi}}, \bibinfo {author} {\bibfnamefont
			{J.}~\bibnamefont {Xiang}}, \bibinfo {author} {\bibfnamefont
			{J.}~\bibnamefont {Finley}}, \bibinfo {author} {\bibfnamefont
			{L.}~\bibnamefont {Fu}}, \bibinfo {author} {\bibfnamefont {R.}~\bibnamefont
			{Cheng}},\ and\ \bibinfo {author} {\bibfnamefont {L.}~\bibnamefont {Liu}},\
	}\bibfield  {title} {\bibinfo {title} {Birefringence-like spin transport via
			linearly polarized antiferromagnetic magnons},\ }\href
	{https://doi.org/10.1038/s41565-020-0703-8} {\bibfield  {journal} {\bibinfo
			{journal} {Nat. Nanotechnol.}\ }\textbf {\bibinfo {volume} {15}},\ \bibinfo
		{pages} {563} (\bibinfo {year} {2020})}\BibitemShut {NoStop}%
	\bibitem [{\citenamefont {Morrish}(1994)}]{1994_Morrish_Book_FEO}%
	\BibitemOpen
	\bibfield  {author} {\bibinfo {author} {\bibfnamefont {A.~H.}\ \bibnamefont
			{Morrish}},\ }\href@noop {} {\emph {\bibinfo {title} {Canted
				antiferromagnetism: hematite}}}\ (\bibinfo  {publisher} {World Scientific},\
	\bibinfo {year} {1994})\BibitemShut {NoStop}%
	\bibitem [{\citenamefont {Pliarev}\ \emph {et~al.}(1969)\citenamefont
		{Pliarev}, \citenamefont {Sinil},\ and\ \citenamefont
		{Smolenskii}}]{1969_Pisarev_MLB_1}%
	\BibitemOpen
	\bibfield  {author} {\bibinfo {author} {\bibfnamefont {R.~V.}\ \bibnamefont
			{Pliarev}}, \bibinfo {author} {\bibfnamefont {I.~G.}\ \bibnamefont {Sinil}},\
		and\ \bibinfo {author} {\bibfnamefont {G.~A.}\ \bibnamefont {Smolenskii}},\
	}\bibfield  {title} {\bibinfo {title} {Quadratic magneto-optic effects in
			ferro- and antiferromagnets},\ }\href
	{http://jetpletters.ru/ps/0/article_25287.shtml} {\bibfield  {journal}
		{\bibinfo  {journal} {ZhETF Pis. Red.}\ }\textbf {\bibinfo {volume} {9}},\
		\bibinfo {pages} {112} (\bibinfo {year} {1969})}\BibitemShut {NoStop}%
	\bibitem [{\citenamefont {Pisarev}\ \emph {et~al.}(1969)\citenamefont
		{Pisarev}, \citenamefont {Sinii},\ and\ \citenamefont
		{Smolenskii}}]{1969_Pisarev_MLB_2}%
	\BibitemOpen
	\bibfield  {author} {\bibinfo {author} {\bibfnamefont {R.~V.}\ \bibnamefont
			{Pisarev}}, \bibinfo {author} {\bibfnamefont {I.~G.}\ \bibnamefont {Sinii}},\
		and\ \bibinfo {author} {\bibfnamefont {G.~A.}\ \bibnamefont {Smolenskii}},\
	}\bibfield  {title} {\bibinfo {title} {Turning of magnetic sublattices and
			anomalies of the {Cotton-Mouton} effect in terbium iron garnet and in
			hematite},\ }\href {http://jetpletters.ru/ps/0/article_25331.shtml}
	{\bibfield  {journal} {\bibinfo  {journal} {ZhETF Pis. Red.}\ }\textbf
		{\bibinfo {volume} {9}},\ \bibinfo {pages} {294} (\bibinfo {year}
		{1969})}\BibitemShut {NoStop}%
	\bibitem [{\citenamefont {{R. V. Pisarev}}(1971)}]{1971_Pisarev_MO_0001}%
	\BibitemOpen
	\bibfield  {author} {\bibinfo {author} {\bibnamefont {{R. V. Pisarev}}},\
	}\bibfield  {title} {\bibinfo {title} {Optical gyrotropy and birefringence in
			magnetic crystals},\ }\href {https://doi.org/10.1051/jphyscol:19711377}
	{\bibfield  {journal} {\bibinfo  {journal} {J. Phys. Colloques}\ }\textbf
		{\bibinfo {volume} {32}},\ \bibinfo {pages} {C1} (\bibinfo {year}
		{1971})}\BibitemShut {NoStop}%
	\bibitem [{\citenamefont {Krinchik}\ \emph {et~al.}(1973)\citenamefont
		{Krinchik}, \citenamefont {Khrebtov}, \citenamefont {Askochenskii},\ and\
		\citenamefont {Zubov}}]{1973_Krinchik_MO_Longitudinal}%
	\BibitemOpen
	\bibfield  {author} {\bibinfo {author} {\bibfnamefont {G.~S.}\ \bibnamefont
			{Krinchik}}, \bibinfo {author} {\bibfnamefont {A.~P.}\ \bibnamefont
			{Khrebtov}}, \bibinfo {author} {\bibfnamefont {A.~A.}\ \bibnamefont
			{Askochenskii}},\ and\ \bibinfo {author} {\bibfnamefont {V.~E.}\ \bibnamefont
			{Zubov}},\ }\bibfield  {title} {\bibinfo {title} {Surface magnetism of
			hematite},\ }\href {http://jetpletters.ru/ps/0/article_23596.shtml}
	{\bibfield  {journal} {\bibinfo  {journal} {ZhETF Pis. Red.}\ }\textbf
		{\bibinfo {volume} {17}},\ \bibinfo {pages} {466} (\bibinfo {year}
		{1973})}\BibitemShut {NoStop}%
	\bibitem [{\citenamefont {Krinchik}\ and\ \citenamefont
		{Zubov}(1974)}]{1974_Krinchik_MO_Sweep}%
	\BibitemOpen
	\bibfield  {author} {\bibinfo {author} {\bibfnamefont {G.~S.}\ \bibnamefont
			{Krinchik}}\ and\ \bibinfo {author} {\bibfnamefont {V.~E.}\ \bibnamefont
			{Zubov}},\ }\bibfield  {title} {\bibinfo {title} {Magneto-optical properties
			of weak ferromagnets},\ }\href
	{http://jetpletters.ru/ps/0/article_27257.shtml} {\bibfield  {journal}
		{\bibinfo  {journal} {ZhETF Pis. Red.}\ }\textbf {\bibinfo {volume} {20}},\
		\bibinfo {pages} {307} (\bibinfo {year} {1974})}\BibitemShut {NoStop}%
	\bibitem [{\citenamefont {Zubov}\ \emph {et~al.}(1981)\citenamefont {Zubov},
		\citenamefont {Krinchik},\ and\ \citenamefont {Lyskov}}]{1981_Zubov_MO_spec}%
	\BibitemOpen
	\bibfield  {author} {\bibinfo {author} {\bibfnamefont {V.}~\bibnamefont
			{Zubov}}, \bibinfo {author} {\bibfnamefont {G.}~\bibnamefont {Krinchik}},\
		and\ \bibinfo {author} {\bibfnamefont {V.}~\bibnamefont {Lyskov}},\
	}\bibfield  {title} {\bibinfo {title} {Magneto-optical properties of
			hematite},\ }\href {http://jetp.ras.ru/cgi-bin/e/index/e/54/4/p789?a=list}
	{\bibfield  {journal} {\bibinfo  {journal} {Zh. Eksper. Teor. Fiz}\ }\textbf
		{\bibinfo {volume} {81}},\ \bibinfo {pages} {1489} (\bibinfo {year}
		{1981})}\BibitemShut {NoStop}%
	\bibitem [{\citenamefont {Ivantsov}\ \emph {et~al.}(2020)\citenamefont
		{Ivantsov}, \citenamefont {Ivanova}, \citenamefont {Zharkov}, \citenamefont
		{Molokeev}, \citenamefont {Krylov}, \citenamefont {Gudim},\ and\
		\citenamefont {Edelman}}]{2020_Ivantsov_MCD}%
	\BibitemOpen
	\bibfield  {author} {\bibinfo {author} {\bibfnamefont {R.}~\bibnamefont
			{Ivantsov}}, \bibinfo {author} {\bibfnamefont {O.}~\bibnamefont {Ivanova}},
		\bibinfo {author} {\bibfnamefont {S.}~\bibnamefont {Zharkov}}, \bibinfo
		{author} {\bibfnamefont {M.}~\bibnamefont {Molokeev}}, \bibinfo {author}
		{\bibfnamefont {A.}~\bibnamefont {Krylov}}, \bibinfo {author} {\bibfnamefont
			{I.}~\bibnamefont {Gudim}},\ and\ \bibinfo {author} {\bibfnamefont
			{I.}~\bibnamefont {Edelman}},\ }\bibfield  {title} {\bibinfo {title}
		{Magnetic circular dichroism in the canted antiferromagnet
			$\alpha$\text{\ensuremath{-}}{$\rm Fe_2O_3$}: Bulk single crystal and
			nanocrystals},\ }\href
	{https://doi.org/https://doi.org/10.1016/j.jmmm.2019.166208} {\bibfield
		{journal} {\bibinfo  {journal} {J. Magn. Magn. Mater.}\ }\textbf {\bibinfo
			{volume} {498}},\ \bibinfo {pages} {166208} (\bibinfo {year}
		{2020})}\BibitemShut {NoStop}%
	\bibitem [{\citenamefont {Appel}\ \emph {et~al.}(1990)\citenamefont {Appel},
		\citenamefont {Hoffmann},\ and\ \citenamefont
		{Soffel}}]{1990_Appel_FEO_domain}%
	\BibitemOpen
	\bibfield  {author} {\bibinfo {author} {\bibfnamefont {E.}~\bibnamefont
			{Appel}}, \bibinfo {author} {\bibfnamefont {V.}~\bibnamefont {Hoffmann}},\
		and\ \bibinfo {author} {\bibfnamefont {H.}~\bibnamefont {Soffel}},\
	}\bibfield  {title} {\bibinfo {title} {Magneto-optical {Kerr} effect in
			(titano)magnetite, pyrrhotite and hematite},\ }\href
	{https://doi.org/https://doi.org/10.1016/0031-9201(90)90073-7} {\bibfield
		{journal} {\bibinfo  {journal} {Phys. Earth Planet In.}\ }\textbf {\bibinfo
			{volume} {65}},\ \bibinfo {pages} {36} (\bibinfo {year} {1990})}\BibitemShut
	{NoStop}%
	\bibitem [{\citenamefont {Kimel}\ \emph {et~al.}(2024)\citenamefont {Kimel},
		\citenamefont {Rasing},\ and\ \citenamefont
		{Ivanov}}]{2024_Kimel_JMMM_alterm_MO}%
	\BibitemOpen
	\bibfield  {author} {\bibinfo {author} {\bibfnamefont {A.}~\bibnamefont
			{Kimel}}, \bibinfo {author} {\bibfnamefont {T.}~\bibnamefont {Rasing}},\ and\
		\bibinfo {author} {\bibfnamefont {B.}~\bibnamefont {Ivanov}},\ }\bibfield
	{title} {\bibinfo {title} {Optical read-out and control of antiferromagnetic
			{Néel} vector in altermagnets and beyond},\ }\href
	{https://doi.org/https://doi.org/10.1016/j.jmmm.2024.172039} {\bibfield
		{journal} {\bibinfo  {journal} {J. Magn. Magn. Mater.}\ }\textbf {\bibinfo
			{volume} {598}},\ \bibinfo {pages} {172039} (\bibinfo {year}
		{2024})}\BibitemShut {NoStop}%
	\bibitem [{\citenamefont {Krichevtsov}\ \emph {et~al.}(1981)\citenamefont
		{Krichevtsov}, \citenamefont {Mukimov}, \citenamefont {Pisarev},\ and\
		\citenamefont {Ruvinshtein}}]{1981_YFeO3}%
	\BibitemOpen
	\bibfield  {author} {\bibinfo {author} {\bibfnamefont {B.~B.}\ \bibnamefont
			{Krichevtsov}}, \bibinfo {author} {\bibfnamefont {K.~M.}\ \bibnamefont
			{Mukimov}}, \bibinfo {author} {\bibfnamefont {R.~V.}\ \bibnamefont
			{Pisarev}},\ and\ \bibinfo {author} {\bibfnamefont {M.~M.}\ \bibnamefont
			{Ruvinshtein}},\ }\bibfield  {title} {\bibinfo {title} {Antiferromagnetic and
			ferromagnetic faraday effect in yttrium orthoferrite {$\rm YFeO_3$}},\ }\href
	{http://jetpletters.ru/ps/0/article_23218.shtml} {\bibfield  {journal}
		{\bibinfo  {journal} {Pis'ma Zh. Eksp. Teor. Fiz.}\ }\textbf {\bibinfo
			{volume} {34}},\ \bibinfo {pages} {399} (\bibinfo {year} {1981})}\BibitemShut
	{NoStop}%
	\bibitem [{\citenamefont {Šmejkal}\ \emph {et~al.}(2022)\citenamefont
		{Šmejkal}, \citenamefont {Sinova},\ and\ \citenamefont
		{Jungwirth}}]{2022_Smejkal_PRX_alterm}%
	\BibitemOpen
	\bibfield  {author} {\bibinfo {author} {\bibfnamefont {L.}~\bibnamefont
			{Šmejkal}}, \bibinfo {author} {\bibfnamefont {J.}~\bibnamefont {Sinova}},\
		and\ \bibinfo {author} {\bibfnamefont {T.}~\bibnamefont {Jungwirth}},\
	}\bibfield  {title} {\bibinfo {title} {Beyond conventional ferromagnetism and
			antiferromagnetism: A phase with nonrelativistic spin and crystal rotation
			symmetry},\ }\href {https://doi.org/10.1103/PhysRevX.12.031042} {\bibfield
		{journal} {\bibinfo  {journal} {Phys. Rev. X}\ }\textbf {\bibinfo {volume}
			{12}},\ \bibinfo {pages} {031042} (\bibinfo {year} {2022})}\BibitemShut
	{NoStop}%
	\bibitem [{\citenamefont {Song}\ \emph {et~al.}(2025)\citenamefont {Song},
		\citenamefont {Bai}, \citenamefont {Zhou}, \citenamefont {Han}, \citenamefont
		{Reichlova}, \citenamefont {Dil}, \citenamefont {Liu}, \citenamefont {Chen},\
		and\ \citenamefont {Pan}}]{2025_Song_NatRev_altermag}%
	\BibitemOpen
	\bibfield  {author} {\bibinfo {author} {\bibfnamefont {C.}~\bibnamefont
			{Song}}, \bibinfo {author} {\bibfnamefont {H.}~\bibnamefont {Bai}}, \bibinfo
		{author} {\bibfnamefont {Z.}~\bibnamefont {Zhou}}, \bibinfo {author}
		{\bibfnamefont {L.}~\bibnamefont {Han}}, \bibinfo {author} {\bibfnamefont
			{H.}~\bibnamefont {Reichlova}}, \bibinfo {author} {\bibfnamefont {J.~H.}\
			\bibnamefont {Dil}}, \bibinfo {author} {\bibfnamefont {J.}~\bibnamefont
			{Liu}}, \bibinfo {author} {\bibfnamefont {X.}~\bibnamefont {Chen}},\ and\
		\bibinfo {author} {\bibfnamefont {F.}~\bibnamefont {Pan}},\ }\bibfield
	{title} {\bibinfo {title} {Altermagnets as a new class of functional
			materials},\ }\href {https://doi.org/10.1038/s41578-025-00779-1} {\bibfield
		{journal} {\bibinfo  {journal} {Nat. Rev. Mater.}\ }\textbf {\bibinfo
			{volume} {10}},\ \bibinfo {pages} {473} (\bibinfo {year} {2025})}\BibitemShut
	{NoStop}%
	\bibitem [{\citenamefont {Šmejkal}\ \emph {et~al.}(2020)\citenamefont
		{Šmejkal}, \citenamefont {González-Hernández}, \citenamefont {Jungwirth},\
		and\ \citenamefont {Sinova}}]{2020_Smejkal_SciAdv_alterm_AHE}%
	\BibitemOpen
	\bibfield  {author} {\bibinfo {author} {\bibfnamefont {L.}~\bibnamefont
			{Šmejkal}}, \bibinfo {author} {\bibfnamefont {R.}~\bibnamefont
			{González-Hernández}}, \bibinfo {author} {\bibfnamefont {T.}~\bibnamefont
			{Jungwirth}},\ and\ \bibinfo {author} {\bibfnamefont {J.}~\bibnamefont
			{Sinova}},\ }\bibfield  {title} {\bibinfo {title} {Crystal time-reversal
			symmetry breaking and spontaneous {Hall} effect in collinear
			antiferromagnets},\ }\href {https://doi.org/10.1126/sciadv.aaz8809}
	{\bibfield  {journal} {\bibinfo  {journal} {Sci. Adv.}\ }\textbf {\bibinfo
			{volume} {6}},\ \bibinfo {pages} {eaaz8809} (\bibinfo {year}
		{2020})}\BibitemShut {NoStop}%
	\bibitem [{\citenamefont {Feng}\ \emph {et~al.}(2022)\citenamefont {Feng},
		\citenamefont {Zhou}, \citenamefont {{\v{S}}mejkal}, \citenamefont {Wu},
		\citenamefont {Zhu}, \citenamefont {Guo}, \citenamefont
		{Gonz{\'a}lez-Hern{\'a}ndez}, \citenamefont {Wang}, \citenamefont {Yan},
		\citenamefont {Qin}, \citenamefont {Zhang}, \citenamefont {Wu}, \citenamefont
		{Chen}, \citenamefont {Meng}, \citenamefont {Liu}, \citenamefont {Xia},
		\citenamefont {Sinova}, \citenamefont {Jungwirth},\ and\ \citenamefont
		{Liu}}]{2022_Feng_NatElec_RuO2_AHE}%
	\BibitemOpen
	\bibfield  {author} {\bibinfo {author} {\bibfnamefont {Z.}~\bibnamefont
			{Feng}}, \bibinfo {author} {\bibfnamefont {X.}~\bibnamefont {Zhou}}, \bibinfo
		{author} {\bibfnamefont {L.}~\bibnamefont {{\v{S}}mejkal}}, \bibinfo {author}
		{\bibfnamefont {L.}~\bibnamefont {Wu}}, \bibinfo {author} {\bibfnamefont
			{Z.}~\bibnamefont {Zhu}}, \bibinfo {author} {\bibfnamefont {H.}~\bibnamefont
			{Guo}}, \bibinfo {author} {\bibfnamefont {R.}~\bibnamefont
			{Gonz{\'a}lez-Hern{\'a}ndez}}, \bibinfo {author} {\bibfnamefont
			{X.}~\bibnamefont {Wang}}, \bibinfo {author} {\bibfnamefont {H.}~\bibnamefont
			{Yan}}, \bibinfo {author} {\bibfnamefont {P.}~\bibnamefont {Qin}}, \bibinfo
		{author} {\bibfnamefont {X.}~\bibnamefont {Zhang}}, \bibinfo {author}
		{\bibfnamefont {H.}~\bibnamefont {Wu}}, \bibinfo {author} {\bibfnamefont
			{H.}~\bibnamefont {Chen}}, \bibinfo {author} {\bibfnamefont {Z.}~\bibnamefont
			{Meng}}, \bibinfo {author} {\bibfnamefont {L.}~\bibnamefont {Liu}}, \bibinfo
		{author} {\bibfnamefont {Z.}~\bibnamefont {Xia}}, \bibinfo {author}
		{\bibfnamefont {J.}~\bibnamefont {Sinova}}, \bibinfo {author} {\bibfnamefont
			{T.}~\bibnamefont {Jungwirth}},\ and\ \bibinfo {author} {\bibfnamefont
			{Z.}~\bibnamefont {Liu}},\ }\bibfield  {title} {\bibinfo {title} {An
			anomalous {Hall} effect in altermagnetic ruthenium dioxide},\ }\href
	{https://doi.org/10.1038/s41928-022-00866-z} {\bibfield  {journal} {\bibinfo
			{journal} {Nat. Electron.}\ }\textbf {\bibinfo {volume} {5}},\ \bibinfo
		{pages} {735} (\bibinfo {year} {2022})}\BibitemShut {NoStop}%
	\bibitem [{\citenamefont {Gonzalez~Betancourt}\ \emph
		{et~al.}(2023)\citenamefont {Gonzalez~Betancourt}, \citenamefont
		{Zub\'a\ifmmode~\check{c}\else \v{c}\fi{}}, \citenamefont
		{Gonzalez-Hernandez}, \citenamefont {Geishendorf}, \citenamefont {\ifmmode
			\check{S}\else \v{S}\fi{}ob\'a\ifmmode~\check{n}\else \v{n}\fi{}},
		\citenamefont {Springholz}, \citenamefont {Olejn\'{\i}k}, \citenamefont
		{\ifmmode~\check{S}\else \v{S}\fi{}mejkal}, \citenamefont {Sinova},
		\citenamefont {Jungwirth}, \citenamefont {Goennenwein}, \citenamefont
		{Thomas}, \citenamefont {Reichlov\'a}, \citenamefont {\ifmmode~\check{Z}\else
			\v{Z}\fi{}elezn\'y},\ and\ \citenamefont
		{Kriegner}}]{2023_Gonzalez_PRL_MnTe_AHE}%
	\BibitemOpen
	\bibfield  {author} {\bibinfo {author} {\bibfnamefont {R.~D.}\ \bibnamefont
			{Gonzalez~Betancourt}}, \bibinfo {author} {\bibfnamefont {J.}~\bibnamefont
			{Zub\'a\ifmmode~\check{c}\else \v{c}\fi{}}}, \bibinfo {author} {\bibfnamefont
			{R.}~\bibnamefont {Gonzalez-Hernandez}}, \bibinfo {author} {\bibfnamefont
			{K.}~\bibnamefont {Geishendorf}}, \bibinfo {author} {\bibfnamefont
			{Z.}~\bibnamefont {\ifmmode \check{S}\else
				\v{S}\fi{}ob\'a\ifmmode~\check{n}\else \v{n}\fi{}}}, \bibinfo {author}
		{\bibfnamefont {G.}~\bibnamefont {Springholz}}, \bibinfo {author}
		{\bibfnamefont {K.}~\bibnamefont {Olejn\'{\i}k}}, \bibinfo {author}
		{\bibfnamefont {L.}~\bibnamefont {\ifmmode~\check{S}\else \v{S}\fi{}mejkal}},
		\bibinfo {author} {\bibfnamefont {J.}~\bibnamefont {Sinova}}, \bibinfo
		{author} {\bibfnamefont {T.}~\bibnamefont {Jungwirth}}, \bibinfo {author}
		{\bibfnamefont {S.~T.~B.}\ \bibnamefont {Goennenwein}}, \bibinfo {author}
		{\bibfnamefont {A.}~\bibnamefont {Thomas}}, \bibinfo {author} {\bibfnamefont
			{H.}~\bibnamefont {Reichlov\'a}}, \bibinfo {author} {\bibfnamefont
			{J.}~\bibnamefont {\ifmmode~\check{Z}\else \v{Z}\fi{}elezn\'y}},\ and\
		\bibinfo {author} {\bibfnamefont {D.}~\bibnamefont {Kriegner}},\ }\bibfield
	{title} {\bibinfo {title} {Spontaneous anomalous {Hall} effect arising from
			an unconventional compensated magnetic phase in a semiconductor},\ }\href
	{https://doi.org/10.1103/PhysRevLett.130.036702} {\bibfield  {journal}
		{\bibinfo  {journal} {Phys. Rev. Lett.}\ }\textbf {\bibinfo {volume} {130}},\
		\bibinfo {pages} {036702} (\bibinfo {year} {2023})}\BibitemShut {NoStop}%
	\bibitem [{\citenamefont {Takagi}\ \emph {et~al.}(2025)\citenamefont {Takagi},
		\citenamefont {Hirakida}, \citenamefont {Settai}, \citenamefont {Oiwa},
		\citenamefont {Takagi}, \citenamefont {Kitaori}, \citenamefont {Yamauchi},
		\citenamefont {Inoue}, \citenamefont {Yamaura}, \citenamefont
		{Nishio-Hamane}, \citenamefont {Itoh}, \citenamefont {Aji}, \citenamefont
		{Saito}, \citenamefont {Nakajima}, \citenamefont {Nomoto}, \citenamefont
		{Arita},\ and\ \citenamefont {Seki}}]{2024_Takagi_NatMat_FeS_AHE}%
	\BibitemOpen
	\bibfield  {author} {\bibinfo {author} {\bibfnamefont {R.}~\bibnamefont
			{Takagi}}, \bibinfo {author} {\bibfnamefont {R.}~\bibnamefont {Hirakida}},
		\bibinfo {author} {\bibfnamefont {Y.}~\bibnamefont {Settai}}, \bibinfo
		{author} {\bibfnamefont {R.}~\bibnamefont {Oiwa}}, \bibinfo {author}
		{\bibfnamefont {H.}~\bibnamefont {Takagi}}, \bibinfo {author} {\bibfnamefont
			{A.}~\bibnamefont {Kitaori}}, \bibinfo {author} {\bibfnamefont
			{K.}~\bibnamefont {Yamauchi}}, \bibinfo {author} {\bibfnamefont
			{H.}~\bibnamefont {Inoue}}, \bibinfo {author} {\bibfnamefont {J.-i.}\
			\bibnamefont {Yamaura}}, \bibinfo {author} {\bibfnamefont {D.}~\bibnamefont
			{Nishio-Hamane}}, \bibinfo {author} {\bibfnamefont {S.}~\bibnamefont {Itoh}},
		\bibinfo {author} {\bibfnamefont {S.}~\bibnamefont {Aji}}, \bibinfo {author}
		{\bibfnamefont {H.}~\bibnamefont {Saito}}, \bibinfo {author} {\bibfnamefont
			{T.}~\bibnamefont {Nakajima}}, \bibinfo {author} {\bibfnamefont
			{T.}~\bibnamefont {Nomoto}}, \bibinfo {author} {\bibfnamefont
			{R.}~\bibnamefont {Arita}},\ and\ \bibinfo {author} {\bibfnamefont
			{S.}~\bibnamefont {Seki}},\ }\bibfield  {title} {\bibinfo {title}
		{Spontaneous {Hall} effect induced by collinear antiferromagnetic order at
			room temperature},\ }\href {https://doi.org/10.1038/s41563-024-02058-w}
	{\bibfield  {journal} {\bibinfo  {journal} {Nat. Mater.}\ }\textbf {\bibinfo
			{volume} {24}},\ \bibinfo {pages} {63} (\bibinfo {year} {2025})}\BibitemShut
	{NoStop}%
	\bibitem [{\citenamefont {Gonz\'alez-Hern\'andez}\ \emph
		{et~al.}(2021)\citenamefont {Gonz\'alez-Hern\'andez}, \citenamefont
		{\ifmmode~\check{S}\else \v{S}\fi{}mejkal}, \citenamefont {V\'yborn\'y},
		\citenamefont {Yahagi}, \citenamefont {Sinova}, \citenamefont {Jungwirth},\
		and\ \citenamefont {\ifmmode~\check{Z}\else
			\v{Z}\fi{}elezn\'y}}]{2021_Gonzalez_PRL_alterm}%
	\BibitemOpen
	\bibfield  {author} {\bibinfo {author} {\bibfnamefont {R.}~\bibnamefont
			{Gonz\'alez-Hern\'andez}}, \bibinfo {author} {\bibfnamefont {L.}~\bibnamefont
			{\ifmmode~\check{S}\else \v{S}\fi{}mejkal}}, \bibinfo {author} {\bibfnamefont
			{K.}~\bibnamefont {V\'yborn\'y}}, \bibinfo {author} {\bibfnamefont
			{Y.}~\bibnamefont {Yahagi}}, \bibinfo {author} {\bibfnamefont
			{J.}~\bibnamefont {Sinova}}, \bibinfo {author} {\bibfnamefont {T.~c.~v.}\
			\bibnamefont {Jungwirth}},\ and\ \bibinfo {author} {\bibfnamefont
			{J.}~\bibnamefont {\ifmmode~\check{Z}\else \v{Z}\fi{}elezn\'y}},\ }\bibfield
	{title} {\bibinfo {title} {Efficient electrical spin splitter based on
			nonrelativistic collinear antiferromagnetism},\ }\href
	{https://doi.org/10.1103/PhysRevLett.126.127701} {\bibfield  {journal}
		{\bibinfo  {journal} {Phys. Rev. Lett.}\ }\textbf {\bibinfo {volume} {126}},\
		\bibinfo {pages} {127701} (\bibinfo {year} {2021})}\BibitemShut {NoStop}%
	\bibitem [{\citenamefont {Bai}\ \emph {et~al.}(2023)\citenamefont {Bai},
		\citenamefont {Zhang}, \citenamefont {Zhou}, \citenamefont {Chen},
		\citenamefont {Wan}, \citenamefont {Han}, \citenamefont {Zhu}, \citenamefont
		{Liang}, \citenamefont {Su}, \citenamefont {Han}, \citenamefont {Pan},\ and\
		\citenamefont {Song}}]{2023_Bai_PRL_RuO2_spin}%
	\BibitemOpen
	\bibfield  {author} {\bibinfo {author} {\bibfnamefont {H.}~\bibnamefont
			{Bai}}, \bibinfo {author} {\bibfnamefont {Y.~C.}\ \bibnamefont {Zhang}},
		\bibinfo {author} {\bibfnamefont {Y.~J.}\ \bibnamefont {Zhou}}, \bibinfo
		{author} {\bibfnamefont {P.}~\bibnamefont {Chen}}, \bibinfo {author}
		{\bibfnamefont {C.~H.}\ \bibnamefont {Wan}}, \bibinfo {author} {\bibfnamefont
			{L.}~\bibnamefont {Han}}, \bibinfo {author} {\bibfnamefont {W.~X.}\
			\bibnamefont {Zhu}}, \bibinfo {author} {\bibfnamefont {S.~X.}\ \bibnamefont
			{Liang}}, \bibinfo {author} {\bibfnamefont {Y.~C.}\ \bibnamefont {Su}},
		\bibinfo {author} {\bibfnamefont {X.~F.}\ \bibnamefont {Han}}, \bibinfo
		{author} {\bibfnamefont {F.}~\bibnamefont {Pan}},\ and\ \bibinfo {author}
		{\bibfnamefont {C.}~\bibnamefont {Song}},\ }\bibfield  {title} {\bibinfo
		{title} {Efficient spin-to-charge conversion via altermagnetic spin splitting
			effect in antiferromagnet {${\mathrm{RuO}}_{2}$}},\ }\href
	{https://doi.org/10.1103/PhysRevLett.130.216701} {\bibfield  {journal}
		{\bibinfo  {journal} {Phys. Rev. Lett.}\ }\textbf {\bibinfo {volume} {130}},\
		\bibinfo {pages} {216701} (\bibinfo {year} {2023})}\BibitemShut {NoStop}%
	\bibitem [{\citenamefont {Lovesey}\ \emph {et~al.}(2023)\citenamefont
		{Lovesey}, \citenamefont {Khalyavin},\ and\ \citenamefont {van~der
			Laan}}]{2023_Lovesey_PRB_MnTe_alterm}%
	\BibitemOpen
	\bibfield  {author} {\bibinfo {author} {\bibfnamefont {S.~W.}\ \bibnamefont
			{Lovesey}}, \bibinfo {author} {\bibfnamefont {D.~D.}\ \bibnamefont
			{Khalyavin}},\ and\ \bibinfo {author} {\bibfnamefont {G.}~\bibnamefont
			{van~der Laan}},\ }\bibfield  {title} {\bibinfo {title} {Templates for
			magnetic symmetry and altermagnetism in hexagonal {MnTe}},\ }\href
	{https://doi.org/10.1103/PhysRevB.108.174437} {\bibfield  {journal} {\bibinfo
			{journal} {Phys. Rev. B}\ }\textbf {\bibinfo {volume} {108}},\ \bibinfo
		{pages} {174437} (\bibinfo {year} {2023})}\BibitemShut {NoStop}%
	\bibitem [{\citenamefont {Hariki}\ \emph {et~al.}(2024)\citenamefont {Hariki},
		\citenamefont {Dal~Din}, \citenamefont {Amin}, \citenamefont {Yamaguchi},
		\citenamefont {Badura}, \citenamefont {Kriegner}, \citenamefont {Edmonds},
		\citenamefont {Campion}, \citenamefont {Wadley}, \citenamefont {Backes},
		\citenamefont {Veiga}, \citenamefont {Dhesi}, \citenamefont {Springholz},
		\citenamefont {\ifmmode~\check{S}\else \v{S}\fi{}mejkal}, \citenamefont
		{V\'yborn\'y}, \citenamefont {Jungwirth},\ and\ \citenamefont
		{Kune\ifmmode~\check{s}\else \v{s}\fi{}}}]{2024_Hariki_PRL_MnTe_XMCD}%
	\BibitemOpen
	\bibfield  {author} {\bibinfo {author} {\bibfnamefont {A.}~\bibnamefont
			{Hariki}}, \bibinfo {author} {\bibfnamefont {A.}~\bibnamefont {Dal~Din}},
		\bibinfo {author} {\bibfnamefont {O.~J.}\ \bibnamefont {Amin}}, \bibinfo
		{author} {\bibfnamefont {T.}~\bibnamefont {Yamaguchi}}, \bibinfo {author}
		{\bibfnamefont {A.}~\bibnamefont {Badura}}, \bibinfo {author} {\bibfnamefont
			{D.}~\bibnamefont {Kriegner}}, \bibinfo {author} {\bibfnamefont {K.~W.}\
			\bibnamefont {Edmonds}}, \bibinfo {author} {\bibfnamefont {R.~P.}\
			\bibnamefont {Campion}}, \bibinfo {author} {\bibfnamefont {P.}~\bibnamefont
			{Wadley}}, \bibinfo {author} {\bibfnamefont {D.}~\bibnamefont {Backes}},
		\bibinfo {author} {\bibfnamefont {L.~S.~I.}\ \bibnamefont {Veiga}}, \bibinfo
		{author} {\bibfnamefont {S.~S.}\ \bibnamefont {Dhesi}}, \bibinfo {author}
		{\bibfnamefont {G.}~\bibnamefont {Springholz}}, \bibinfo {author}
		{\bibfnamefont {L.}~\bibnamefont {\ifmmode~\check{S}\else \v{S}\fi{}mejkal}},
		\bibinfo {author} {\bibfnamefont {K.}~\bibnamefont {V\'yborn\'y}}, \bibinfo
		{author} {\bibfnamefont {T.}~\bibnamefont {Jungwirth}},\ and\ \bibinfo
		{author} {\bibfnamefont {J.}~\bibnamefont {Kune\ifmmode~\check{s}\else
				\v{s}\fi{}}},\ }\bibfield  {title} {\bibinfo {title} {X-ray magnetic circular
			dichroism in altermagnetic $\ensuremath{\alpha}$-{MnTe}},\ }\href
	{https://doi.org/10.1103/PhysRevLett.132.176701} {\bibfield  {journal}
		{\bibinfo  {journal} {Phys. Rev. Lett.}\ }\textbf {\bibinfo {volume} {132}},\
		\bibinfo {pages} {176701} (\bibinfo {year} {2024})}\BibitemShut {NoStop}%
	\bibitem [{\citenamefont {Amin}\ \emph {et~al.}(2024)\citenamefont {Amin},
		\citenamefont {Dal~Din}, \citenamefont {Golias}, \citenamefont {Niu},
		\citenamefont {Zakharov}, \citenamefont {Fromage}, \citenamefont {Fields},
		\citenamefont {Heywood}, \citenamefont {Cousins}, \citenamefont
		{Maccherozzi}, \citenamefont {Krempask{\'y}}, \citenamefont {Dil},
		\citenamefont {Kriegner}, \citenamefont {Kiraly}, \citenamefont {Campion},
		\citenamefont {Rushforth}, \citenamefont {Edmonds}, \citenamefont {Dhesi},
		\citenamefont {{\v{S}}mejkal}, \citenamefont {Jungwirth},\ and\ \citenamefont
		{Wadley}}]{2024_Amin_Nat_MnTe_mapping}%
	\BibitemOpen
	\bibfield  {author} {\bibinfo {author} {\bibfnamefont {O.~J.}\ \bibnamefont
			{Amin}}, \bibinfo {author} {\bibfnamefont {A.}~\bibnamefont {Dal~Din}},
		\bibinfo {author} {\bibfnamefont {E.}~\bibnamefont {Golias}}, \bibinfo
		{author} {\bibfnamefont {Y.}~\bibnamefont {Niu}}, \bibinfo {author}
		{\bibfnamefont {A.}~\bibnamefont {Zakharov}}, \bibinfo {author}
		{\bibfnamefont {S.~C.}\ \bibnamefont {Fromage}}, \bibinfo {author}
		{\bibfnamefont {C.~J.~B.}\ \bibnamefont {Fields}}, \bibinfo {author}
		{\bibfnamefont {S.~L.}\ \bibnamefont {Heywood}}, \bibinfo {author}
		{\bibfnamefont {R.~B.}\ \bibnamefont {Cousins}}, \bibinfo {author}
		{\bibfnamefont {F.}~\bibnamefont {Maccherozzi}}, \bibinfo {author}
		{\bibfnamefont {J.}~\bibnamefont {Krempask{\'y}}}, \bibinfo {author}
		{\bibfnamefont {J.~H.}\ \bibnamefont {Dil}}, \bibinfo {author} {\bibfnamefont
			{D.}~\bibnamefont {Kriegner}}, \bibinfo {author} {\bibfnamefont
			{B.}~\bibnamefont {Kiraly}}, \bibinfo {author} {\bibfnamefont {R.~P.}\
			\bibnamefont {Campion}}, \bibinfo {author} {\bibfnamefont {A.~W.}\
			\bibnamefont {Rushforth}}, \bibinfo {author} {\bibfnamefont {K.~W.}\
			\bibnamefont {Edmonds}}, \bibinfo {author} {\bibfnamefont {S.~S.}\
			\bibnamefont {Dhesi}}, \bibinfo {author} {\bibfnamefont {L.}~\bibnamefont
			{{\v{S}}mejkal}}, \bibinfo {author} {\bibfnamefont {T.}~\bibnamefont
			{Jungwirth}},\ and\ \bibinfo {author} {\bibfnamefont {P.}~\bibnamefont
			{Wadley}},\ }\bibfield  {title} {\bibinfo {title} {Nanoscale imaging and
			control of altermagnetism in {MnTe}},\ }\href
	{https://doi.org/10.1038/s41586-024-08234-x} {\bibfield  {journal} {\bibinfo
			{journal} {Nature}\ }\textbf {\bibinfo {volume} {636}},\ \bibinfo {pages}
		{348} (\bibinfo {year} {2024})}\BibitemShut {NoStop}%
	\bibitem [{\citenamefont {Yao}\ \emph {et~al.}(2004)\citenamefont {Yao},
		\citenamefont {Kleinman}, \citenamefont {MacDonald}, \citenamefont {Sinova},
		\citenamefont {Jungwirth}, \citenamefont {Wang}, \citenamefont {Wang},\ and\
		\citenamefont {Niu}}]{2004_Yao}%
	\BibitemOpen
	\bibfield  {author} {\bibinfo {author} {\bibfnamefont {Y.}~\bibnamefont
			{Yao}}, \bibinfo {author} {\bibfnamefont {L.}~\bibnamefont {Kleinman}},
		\bibinfo {author} {\bibfnamefont {A.~H.}\ \bibnamefont {MacDonald}}, \bibinfo
		{author} {\bibfnamefont {J.}~\bibnamefont {Sinova}}, \bibinfo {author}
		{\bibfnamefont {T.}~\bibnamefont {Jungwirth}}, \bibinfo {author}
		{\bibfnamefont {D.-s.}\ \bibnamefont {Wang}}, \bibinfo {author}
		{\bibfnamefont {E.}~\bibnamefont {Wang}},\ and\ \bibinfo {author}
		{\bibfnamefont {Q.}~\bibnamefont {Niu}},\ }\bibfield  {title} {\bibinfo
		{title} {First principles calculation of anomalous {Hall} conductivity in
			ferromagnetic bcc {Fe}},\ }\href
	{https://doi.org/10.1103/PhysRevLett.92.037204} {\bibfield  {journal}
		{\bibinfo  {journal} {Phys. Rev. Lett.}\ }\textbf {\bibinfo {volume} {92}},\
		\bibinfo {pages} {037204} (\bibinfo {year} {2004})}\BibitemShut {NoStop}%
	\bibitem [{\citenamefont {Verbeek}\ \emph {et~al.}(2024)\citenamefont
		{Verbeek}, \citenamefont {Voderholzer}, \citenamefont {Sch\"aren},
		\citenamefont {Gachnang}, \citenamefont {Spaldin},\ and\ \citenamefont
		{Bhowal}}]{2024_Verbeek_PRR_FEO_alterm}%
	\BibitemOpen
	\bibfield  {author} {\bibinfo {author} {\bibfnamefont {X.~H.}\ \bibnamefont
			{Verbeek}}, \bibinfo {author} {\bibfnamefont {D.}~\bibnamefont
			{Voderholzer}}, \bibinfo {author} {\bibfnamefont {S.}~\bibnamefont
			{Sch\"aren}}, \bibinfo {author} {\bibfnamefont {Y.}~\bibnamefont {Gachnang}},
		\bibinfo {author} {\bibfnamefont {N.~A.}\ \bibnamefont {Spaldin}},\ and\
		\bibinfo {author} {\bibfnamefont {S.}~\bibnamefont {Bhowal}},\ }\bibfield
	{title} {\bibinfo {title} {Nonrelativistic ferromagnetotriakontadipolar order
			and spin splitting in hematite},\ }\href
	{https://doi.org/10.1103/PhysRevResearch.6.043157} {\bibfield  {journal}
		{\bibinfo  {journal} {Phys. Rev. Res.}\ }\textbf {\bibinfo {volume} {6}},\
		\bibinfo {pages} {043157} (\bibinfo {year} {2024})}\BibitemShut {NoStop}%
	\bibitem [{\citenamefont {Galindez-Ruales}\ \emph {et~al.}(2025)\citenamefont
		{Galindez-Ruales}, \citenamefont {Gonzalez-Hernandez}, \citenamefont
		{Schmitt}, \citenamefont {Das}, \citenamefont {Fuhrmann}, \citenamefont
		{Ross}, \citenamefont {Golias}, \citenamefont {Akashdeep}, \citenamefont
		{L{\"u}nenb{\"u}rger}, \citenamefont {Baek}, \citenamefont {Yang},
		\citenamefont {{\v{S}}mejkal}, \citenamefont {Krishna}, \citenamefont
		{Jaeschke-Ubiergo}, \citenamefont {Sinova}, \citenamefont {Rothschild},
		\citenamefont {You}, \citenamefont {Jakob},\ and\ \citenamefont
		{Kl{\"a}ui}}]{2025_Galindez_AM}%
	\BibitemOpen
	\bibfield  {author} {\bibinfo {author} {\bibfnamefont {E.}~\bibnamefont
			{Galindez-Ruales}}, \bibinfo {author} {\bibfnamefont {R.}~\bibnamefont
			{Gonzalez-Hernandez}}, \bibinfo {author} {\bibfnamefont {C.}~\bibnamefont
			{Schmitt}}, \bibinfo {author} {\bibfnamefont {S.}~\bibnamefont {Das}},
		\bibinfo {author} {\bibfnamefont {F.}~\bibnamefont {Fuhrmann}}, \bibinfo
		{author} {\bibfnamefont {A.}~\bibnamefont {Ross}}, \bibinfo {author}
		{\bibfnamefont {E.}~\bibnamefont {Golias}}, \bibinfo {author} {\bibfnamefont
			{A.}~\bibnamefont {Akashdeep}}, \bibinfo {author} {\bibfnamefont
			{L.}~\bibnamefont {L{\"u}nenb{\"u}rger}}, \bibinfo {author} {\bibfnamefont
			{E.}~\bibnamefont {Baek}}, \bibinfo {author} {\bibfnamefont {W.}~\bibnamefont
			{Yang}}, \bibinfo {author} {\bibfnamefont {L.}~\bibnamefont {{\v{S}}mejkal}},
		\bibinfo {author} {\bibfnamefont {V.}~\bibnamefont {Krishna}}, \bibinfo
		{author} {\bibfnamefont {R.}~\bibnamefont {Jaeschke-Ubiergo}}, \bibinfo
		{author} {\bibfnamefont {J.}~\bibnamefont {Sinova}}, \bibinfo {author}
		{\bibfnamefont {A.}~\bibnamefont {Rothschild}}, \bibinfo {author}
		{\bibfnamefont {C.-Y.}\ \bibnamefont {You}}, \bibinfo {author} {\bibfnamefont
			{G.}~\bibnamefont {Jakob}},\ and\ \bibinfo {author} {\bibfnamefont
			{M.}~\bibnamefont {Kl{\"a}ui}},\ }\bibfield  {title} {\bibinfo {title}
		{Revealing the altermagnetism in hematite via {XMCD} imaging and anomalous
			{Hall} electrical transport},\ }\href
	{https://doi.org/10.1002/adma.202505019} {\bibfield  {journal} {\bibinfo
			{journal} {Adv. Mater.}\ }\textbf {\bibinfo {volume} {n/a}},\ \bibinfo
		{pages} {e05019} (\bibinfo {year} {2025})}\BibitemShut {NoStop}%
	\bibitem [{\citenamefont {Kanj}\ \emph {et~al.}(2023)\citenamefont {Kanj},
		\citenamefont {Gomonay}, \citenamefont {Boventer}, \citenamefont
		{Bortolotti}, \citenamefont {Cros}, \citenamefont {Anane},\ and\
		\citenamefont {Lebrun}}]{2023_Kanj_SciAdv_magnon}%
	\BibitemOpen
	\bibfield  {author} {\bibinfo {author} {\bibfnamefont {A.~E.}\ \bibnamefont
			{Kanj}}, \bibinfo {author} {\bibfnamefont {O.}~\bibnamefont {Gomonay}},
		\bibinfo {author} {\bibfnamefont {I.}~\bibnamefont {Boventer}}, \bibinfo
		{author} {\bibfnamefont {P.}~\bibnamefont {Bortolotti}}, \bibinfo {author}
		{\bibfnamefont {V.}~\bibnamefont {Cros}}, \bibinfo {author} {\bibfnamefont
			{A.}~\bibnamefont {Anane}},\ and\ \bibinfo {author} {\bibfnamefont
			{R.}~\bibnamefont {Lebrun}},\ }\bibfield  {title} {\bibinfo {title}
		{Antiferromagnetic magnon spintronic based on nonreciprocal and
			nondegenerated ultra-fast spin-waves in the canted antiferromagnet
			$\ensuremath{\alpha}$\text{\ensuremath{-}}{$\rm Fe_2O_3$}},\ }\href
	{https://doi.org/10.1126/sciadv.adh1601} {\bibfield  {journal} {\bibinfo
			{journal} {Sci. Adv.}\ }\textbf {\bibinfo {volume} {9}},\ \bibinfo {pages}
		{eadh1601} (\bibinfo {year} {2023})}\BibitemShut {NoStop}%
	\bibitem [{\citenamefont {Chen}\ \emph {et~al.}(2025)\citenamefont {Chen},
		\citenamefont {Jin}, \citenamefont {Yuan}, \citenamefont {Wang},
		\citenamefont {Jia}, \citenamefont {Wei}, \citenamefont {Sheng},
		\citenamefont {Wang}, \citenamefont {Zhang}, \citenamefont {Liu},
		\citenamefont {Yu}, \citenamefont {Ansermet}, \citenamefont {Yan},\ and\
		\citenamefont {Yu}}]{2025_Chen_PRL_FEO_magnon}%
	\BibitemOpen
	\bibfield  {author} {\bibinfo {author} {\bibfnamefont {J.}~\bibnamefont
			{Chen}}, \bibinfo {author} {\bibfnamefont {Z.}~\bibnamefont {Jin}}, \bibinfo
		{author} {\bibfnamefont {R.}~\bibnamefont {Yuan}}, \bibinfo {author}
		{\bibfnamefont {H.}~\bibnamefont {Wang}}, \bibinfo {author} {\bibfnamefont
			{H.}~\bibnamefont {Jia}}, \bibinfo {author} {\bibfnamefont {W.}~\bibnamefont
			{Wei}}, \bibinfo {author} {\bibfnamefont {L.}~\bibnamefont {Sheng}}, \bibinfo
		{author} {\bibfnamefont {J.}~\bibnamefont {Wang}}, \bibinfo {author}
		{\bibfnamefont {Y.}~\bibnamefont {Zhang}}, \bibinfo {author} {\bibfnamefont
			{S.}~\bibnamefont {Liu}}, \bibinfo {author} {\bibfnamefont {D.}~\bibnamefont
			{Yu}}, \bibinfo {author} {\bibfnamefont {J.-P.}\ \bibnamefont {Ansermet}},
		\bibinfo {author} {\bibfnamefont {P.}~\bibnamefont {Yan}},\ and\ \bibinfo
		{author} {\bibfnamefont {H.}~\bibnamefont {Yu}},\ }\bibfield  {title}
	{\bibinfo {title} {Observation of coherent gapless magnons in an
			antiferromagnet},\ }\href {https://doi.org/10.1103/PhysRevLett.134.056701}
	{\bibfield  {journal} {\bibinfo  {journal} {Phys. Rev. Lett.}\ }\textbf
		{\bibinfo {volume} {134}},\ \bibinfo {pages} {056701} (\bibinfo {year}
		{2025})}\BibitemShut {NoStop}%
	\bibitem [{\citenamefont {Xiao}\ \emph {et~al.}(2025)\citenamefont {Xiao},
		\citenamefont {Li}, \citenamefont {Han}, \citenamefont {Gan}, \citenamefont
		{Yang}, \citenamefont {Shao}, \citenamefont {Zhang}, \citenamefont {Gao},
		\citenamefont {Tian},\ and\ \citenamefont {Zhou}}]{2024_Xiao_altermag}%
	\BibitemOpen
	\bibfield  {author} {\bibinfo {author} {\bibfnamefont {R.-C.}\ \bibnamefont
			{Xiao}}, \bibinfo {author} {\bibfnamefont {H.}~\bibnamefont {Li}}, \bibinfo
		{author} {\bibfnamefont {H.}~\bibnamefont {Han}}, \bibinfo {author}
		{\bibfnamefont {W.}~\bibnamefont {Gan}}, \bibinfo {author} {\bibfnamefont
			{M.}~\bibnamefont {Yang}}, \bibinfo {author} {\bibfnamefont {D.-F.}\
			\bibnamefont {Shao}}, \bibinfo {author} {\bibfnamefont {S.-H.}\ \bibnamefont
			{Zhang}}, \bibinfo {author} {\bibfnamefont {Y.}~\bibnamefont {Gao}}, \bibinfo
		{author} {\bibfnamefont {M.}~\bibnamefont {Tian}},\ and\ \bibinfo {author}
		{\bibfnamefont {J.}~\bibnamefont {Zhou}},\ }\bibfield  {title} {\bibinfo
		{title} {Anomalous-hall n{\'e}el textures in altermagnetic materials},\
	}\href {https://doi.org/10.1007/s11433-025-2769-6} {\bibfield  {journal}
		{\bibinfo  {journal} {Sci. China- Phys. Mech. Astron.}\ }\textbf {\bibinfo
			{volume} {69}},\ \bibinfo {pages} {217511} (\bibinfo {year}
		{2025})}\BibitemShut {NoStop}%
	\bibitem [{\citenamefont {Hill}\ \emph {et~al.}(2008)\citenamefont {Hill},
		\citenamefont {Jiao}, \citenamefont {Bruce}, \citenamefont {Harrison},
		\citenamefont {Kockelmann},\ and\ \citenamefont {Ritter}}]{2008_Hill_FEO_ND}%
	\BibitemOpen
	\bibfield  {author} {\bibinfo {author} {\bibfnamefont {A.~H.}\ \bibnamefont
			{Hill}}, \bibinfo {author} {\bibfnamefont {F.}~\bibnamefont {Jiao}}, \bibinfo
		{author} {\bibfnamefont {P.~G.}\ \bibnamefont {Bruce}}, \bibinfo {author}
		{\bibfnamefont {A.}~\bibnamefont {Harrison}}, \bibinfo {author}
		{\bibfnamefont {W.}~\bibnamefont {Kockelmann}},\ and\ \bibinfo {author}
		{\bibfnamefont {C.}~\bibnamefont {Ritter}},\ }\bibfield  {title} {\bibinfo
		{title} {Neutron diffraction study of mesoporous and bulk hematite,
			{$\alpha$-Fe2O3}},\ }\href {https://doi.org/10.1021/cm800009s} {\bibfield
		{journal} {\bibinfo  {journal} {Chem. Mater.}\ }\textbf {\bibinfo {volume}
			{20}},\ \bibinfo {pages} {4891} (\bibinfo {year} {2008})}\BibitemShut
	{NoStop}%
	\bibitem [{\citenamefont {Roig}\ \emph {et~al.}(2025)\citenamefont {Roig},
		\citenamefont {Yu}, \citenamefont {Ekman}, \citenamefont {Kreisel},
		\citenamefont {Andersen},\ and\ \citenamefont {Agterberg}}]{2025_Roig_PRL}%
	\BibitemOpen
	\bibfield  {author} {\bibinfo {author} {\bibfnamefont {M.}~\bibnamefont
			{Roig}}, \bibinfo {author} {\bibfnamefont {Y.}~\bibnamefont {Yu}}, \bibinfo
		{author} {\bibfnamefont {R.~C.}\ \bibnamefont {Ekman}}, \bibinfo {author}
		{\bibfnamefont {A.}~\bibnamefont {Kreisel}}, \bibinfo {author} {\bibfnamefont
			{B.~M.}\ \bibnamefont {Andersen}},\ and\ \bibinfo {author} {\bibfnamefont
			{D.~F.}\ \bibnamefont {Agterberg}},\ }\bibfield  {title} {\bibinfo {title}
		{Quasisymmetry-constrained spin ferromagnetism in altermagnets},\ }\href
	{https://doi.org/10.1103/839n-rckn} {\bibfield  {journal} {\bibinfo
			{journal} {Phys. Rev. Lett.}\ }\textbf {\bibinfo {volume} {135}},\ \bibinfo
		{pages} {016703} (\bibinfo {year} {2025})}\BibitemShut {NoStop}%
	\bibitem [{sup()}]{supp}%
	\BibitemOpen
	\href@noop {} {}\bibinfo {note} {See Supplemental Material at [URL] for the
		XRD data of the experiments, first-principles calculations and formula
		derivation.}\BibitemShut {Stop}%
	\bibitem [{\citenamefont {Querry}(1985)}]{1985_Querry_N}%
	\BibitemOpen
	\bibfield  {author} {\bibinfo {author} {\bibfnamefont {M.~R.}\ \bibnamefont
			{Querry}},\ }\href@noop {} {\emph {\bibinfo {title} {Optical constants}}},\
	\bibinfo {type} {Tech. Rep.}\ (\bibinfo {year} {1985})\BibitemShut {NoStop}%
	\bibitem [{\citenamefont {Johnson}\ and\ \citenamefont
		{Christy}(1972)}]{1972_Johnson_N}%
	\BibitemOpen
	\bibfield  {author} {\bibinfo {author} {\bibfnamefont {P.~B.}\ \bibnamefont
			{Johnson}}\ and\ \bibinfo {author} {\bibfnamefont {R.~W.}\ \bibnamefont
			{Christy}},\ }\bibfield  {title} {\bibinfo {title} {Optical constants of the
			noble metals},\ }\href {https://doi.org/10.1103/PhysRevB.6.4370} {\bibfield
		{journal} {\bibinfo  {journal} {Phys. Rev. B}\ }\textbf {\bibinfo {volume}
			{6}},\ \bibinfo {pages} {4370} (\bibinfo {year} {1972})}\BibitemShut
	{NoStop}%
	\bibitem [{\citenamefont {You}\ and\ \citenamefont
		{Shin}(1998)}]{1998_You_formula}%
	\BibitemOpen
	\bibfield  {author} {\bibinfo {author} {\bibfnamefont {C.-Y.}\ \bibnamefont
			{You}}\ and\ \bibinfo {author} {\bibfnamefont {S.-C.}\ \bibnamefont {Shin}},\
	}\bibfield  {title} {\bibinfo {title} {Generalized analytic formulae for
			magneto-optical {Kerr} effects},\ }\href {https://doi.org/10.1063/1.368058}
	{\bibfield  {journal} {\bibinfo  {journal} {J. Appl. Phys.}\ }\textbf
		{\bibinfo {volume} {84}},\ \bibinfo {pages} {541} (\bibinfo {year}
		{1998})}\BibitemShut {NoStop}%
	\bibitem [{\citenamefont {Qiu}\ and\ \citenamefont
		{Bader}(2000)}]{2000_Qiu_MOreview}%
	\BibitemOpen
	\bibfield  {author} {\bibinfo {author} {\bibfnamefont {Z.~Q.}\ \bibnamefont
			{Qiu}}\ and\ \bibinfo {author} {\bibfnamefont {S.~D.}\ \bibnamefont
			{Bader}},\ }\bibfield  {title} {\bibinfo {title} {Surface magneto-optic
			{Kerr} effect},\ }\href {https://doi.org/10.1063/1.1150496} {\bibfield
		{journal} {\bibinfo  {journal} {Rev. Sci. Instrum.}\ }\textbf {\bibinfo
			{volume} {71}},\ \bibinfo {pages} {1243} (\bibinfo {year}
		{2000})}\BibitemShut {NoStop}%
	\bibitem [{\citenamefont {Higo}\ \emph {et~al.}(2018)\citenamefont {Higo},
		\citenamefont {Man}, \citenamefont {Gopman}, \citenamefont {Wu},
		\citenamefont {Koretsune}, \citenamefont {van~'t Erve}, \citenamefont
		{Kabanov}, \citenamefont {Rees}, \citenamefont {Li}, \citenamefont {Suzuki},
		\citenamefont {Patankar}, \citenamefont {Ikhlas}, \citenamefont {Chien},
		\citenamefont {Arita}, \citenamefont {Shull}, \citenamefont {Orenstein},\
		and\ \citenamefont {Nakatsuji}}]{2018_Higo_MOKE}%
	\BibitemOpen
	\bibfield  {author} {\bibinfo {author} {\bibfnamefont {T.}~\bibnamefont
			{Higo}}, \bibinfo {author} {\bibfnamefont {H.}~\bibnamefont {Man}}, \bibinfo
		{author} {\bibfnamefont {D.~B.}\ \bibnamefont {Gopman}}, \bibinfo {author}
		{\bibfnamefont {L.}~\bibnamefont {Wu}}, \bibinfo {author} {\bibfnamefont
			{T.}~\bibnamefont {Koretsune}}, \bibinfo {author} {\bibfnamefont {O.~M.~J.}\
			\bibnamefont {van~'t Erve}}, \bibinfo {author} {\bibfnamefont {Y.~P.}\
			\bibnamefont {Kabanov}}, \bibinfo {author} {\bibfnamefont {D.}~\bibnamefont
			{Rees}}, \bibinfo {author} {\bibfnamefont {Y.}~\bibnamefont {Li}}, \bibinfo
		{author} {\bibfnamefont {M.-T.}\ \bibnamefont {Suzuki}}, \bibinfo {author}
		{\bibfnamefont {S.}~\bibnamefont {Patankar}}, \bibinfo {author}
		{\bibfnamefont {M.}~\bibnamefont {Ikhlas}}, \bibinfo {author} {\bibfnamefont
			{C.~L.}\ \bibnamefont {Chien}}, \bibinfo {author} {\bibfnamefont
			{R.}~\bibnamefont {Arita}}, \bibinfo {author} {\bibfnamefont {R.~D.}\
			\bibnamefont {Shull}}, \bibinfo {author} {\bibfnamefont {J.}~\bibnamefont
			{Orenstein}},\ and\ \bibinfo {author} {\bibfnamefont {S.}~\bibnamefont
			{Nakatsuji}},\ }\bibfield  {title} {\bibinfo {title} {Large magneto-optical
			{Kerr} effect and imaging of magnetic octupole domains in an
			antiferromagnetic metal},\ }\href {https://doi.org/10.1038/s41566-017-0086-z}
	{\bibfield  {journal} {\bibinfo  {journal} {Nat. Photonics}\ }\textbf
		{\bibinfo {volume} {12}},\ \bibinfo {pages} {73} (\bibinfo {year}
		{2018})}\BibitemShut {NoStop}%
	\bibitem [{\citenamefont {Ebert}(1996)}]{1996_Ebert_RPP}%
	\BibitemOpen
	\bibfield  {author} {\bibinfo {author} {\bibfnamefont {H.}~\bibnamefont
			{Ebert}},\ }\bibfield  {title} {\bibinfo {title} {Magneto-optical effects in
			transition metal systems},\ }\href
	{https://doi.org/10.1088/0034-4885/59/12/003} {\bibfield  {journal} {\bibinfo
			{journal} {Rep. Prog. Phys.}\ }\textbf {\bibinfo {volume} {59}},\ \bibinfo
		{pages} {1665} (\bibinfo {year} {1996})}\BibitemShut {NoStop}%
	\bibitem [{\citenamefont {Sivadas}\ \emph {et~al.}(2016)\citenamefont
		{Sivadas}, \citenamefont {Okamoto},\ and\ \citenamefont
		{Xiao}}]{2016_Xiao_PRL}%
	\BibitemOpen
	\bibfield  {author} {\bibinfo {author} {\bibfnamefont {N.}~\bibnamefont
			{Sivadas}}, \bibinfo {author} {\bibfnamefont {S.}~\bibnamefont {Okamoto}},\
		and\ \bibinfo {author} {\bibfnamefont {D.}~\bibnamefont {Xiao}},\ }\bibfield
	{title} {\bibinfo {title} {Gate-controllable magneto-optic {Kerr} effect in
			layered collinear antiferromagnets},\ }\href
	{https://doi.org/10.1103/PhysRevLett.117.267203} {\bibfield  {journal}
		{\bibinfo  {journal} {Phys. Rev. Lett.}\ }\textbf {\bibinfo {volume} {117}},\
		\bibinfo {pages} {267203} (\bibinfo {year} {2016})}\BibitemShut {NoStop}%
	\bibitem [{\citenamefont {Mostofi}\ \emph {et~al.}(2008)\citenamefont
		{Mostofi}, \citenamefont {Yates}, \citenamefont {Lee}, \citenamefont {Souza},
		\citenamefont {Vanderbilt},\ and\ \citenamefont
		{Marzari}}]{2008_Mostofi_CPC}%
	\BibitemOpen
	\bibfield  {author} {\bibinfo {author} {\bibfnamefont {A.~A.}\ \bibnamefont
			{Mostofi}}, \bibinfo {author} {\bibfnamefont {J.~R.}\ \bibnamefont {Yates}},
		\bibinfo {author} {\bibfnamefont {Y.-S.}\ \bibnamefont {Lee}}, \bibinfo
		{author} {\bibfnamefont {I.}~\bibnamefont {Souza}}, \bibinfo {author}
		{\bibfnamefont {D.}~\bibnamefont {Vanderbilt}},\ and\ \bibinfo {author}
		{\bibfnamefont {N.}~\bibnamefont {Marzari}},\ }\bibfield  {title} {\bibinfo
		{title} {wannier90: A tool for obtaining maximally-localised {Wannier}
			functions},\ }\href
	{https://doi.org/https://doi.org/10.1016/j.cpc.2007.11.016} {\bibfield
		{journal} {\bibinfo  {journal} {Comput. Phys. Commun.}\ }\textbf {\bibinfo
			{volume} {178}},\ \bibinfo {pages} {685} (\bibinfo {year}
		{2008})}\BibitemShut {NoStop}%
	\bibitem [{\citenamefont {Kresse}\ and\ \citenamefont
		{Furthm\"uller}(1996)}]{1996_Kresse_PRB}%
	\BibitemOpen
	\bibfield  {author} {\bibinfo {author} {\bibfnamefont {G.}~\bibnamefont
			{Kresse}}\ and\ \bibinfo {author} {\bibfnamefont {J.}~\bibnamefont
			{Furthm\"uller}},\ }\bibfield  {title} {\bibinfo {title} {Efficient iterative
			schemes for ab initio total-energy calculations using a plane-wave basis
			set},\ }\href {https://doi.org/10.1103/PhysRevB.54.11169} {\bibfield
		{journal} {\bibinfo  {journal} {Phys. Rev. B}\ }\textbf {\bibinfo {volume}
			{54}},\ \bibinfo {pages} {11169} (\bibinfo {year} {1996})}\BibitemShut
	{NoStop}%
	\bibitem [{\citenamefont {Kresse}\ and\ \citenamefont
		{Joubert}(1999)}]{1999_Kresse_PRB}%
	\BibitemOpen
	\bibfield  {author} {\bibinfo {author} {\bibfnamefont {G.}~\bibnamefont
			{Kresse}}\ and\ \bibinfo {author} {\bibfnamefont {D.}~\bibnamefont
			{Joubert}},\ }\bibfield  {title} {\bibinfo {title} {From ultrasoft
			pseudopotentials to the projector augmented-wave method},\ }\href
	{https://doi.org/10.1103/PhysRevB.59.1758} {\bibfield  {journal} {\bibinfo
			{journal} {Phys. Rev. B}\ }\textbf {\bibinfo {volume} {59}},\ \bibinfo
		{pages} {1758} (\bibinfo {year} {1999})}\BibitemShut {NoStop}%
	\bibitem [{\citenamefont {Marder}(2010)}]{2010_Book_Condensed}%
	\BibitemOpen
	\bibfield  {author} {\bibinfo {author} {\bibfnamefont {M.~P.}\ \bibnamefont
			{Marder}},\ }\href@noop {} {\emph {\bibinfo {title} {Condensed Matter
				Physics}}}\ (\bibinfo  {publisher} {John Wiley \& Sons},\ \bibinfo {year}
	{2010})\BibitemShut {NoStop}%
	\bibitem [{\citenamefont {Xiao}\ \emph {et~al.}(2026)\citenamefont {Xiao},
		\citenamefont {Jin}, \citenamefont {Zhang}, \citenamefont {Feng},
		\citenamefont {Shao},\ and\ \citenamefont {Tian}}]{2026_Xiao}%
	\BibitemOpen
	\bibfield  {author} {\bibinfo {author} {\bibfnamefont {R.-C.}\ \bibnamefont
			{Xiao}}, \bibinfo {author} {\bibfnamefont {Y.}~\bibnamefont {Jin}}, \bibinfo
		{author} {\bibfnamefont {Z.-F.}\ \bibnamefont {Zhang}}, \bibinfo {author}
		{\bibfnamefont {Z.-H.}\ \bibnamefont {Feng}}, \bibinfo {author}
		{\bibfnamefont {D.-F.}\ \bibnamefont {Shao}},\ and\ \bibinfo {author}
		{\bibfnamefont {M.}~\bibnamefont {Tian}},\ }\bibfield  {title} {\bibinfo
		{title} {Tensorsymmetry: a package to get symmetry-adapted tensors
			disentangling spin-orbit coupling effect and establishing analytical
			relationship with magnetic order},\ }\href
	{https://doi.org/https://doi.org/10.1016/j.cpc.2025.109872} {\bibfield
		{journal} {\bibinfo  {journal} {Comput. Phys. Commun.}\ }\textbf {\bibinfo
			{volume} {318}},\ \bibinfo {pages} {109872} (\bibinfo {year}
		{2026})}\BibitemShut {NoStop}%
	\bibitem [{\citenamefont {Liu}\ \emph {et~al.}(2025)\citenamefont {Liu},
		\citenamefont {Wei}, \citenamefont {Peng}, \citenamefont {Hou}, \citenamefont
		{Gao},\ and\ \citenamefont {Niu}}]{2025_Liu_PRX}%
	\BibitemOpen
	\bibfield  {author} {\bibinfo {author} {\bibfnamefont {Z.}~\bibnamefont
			{Liu}}, \bibinfo {author} {\bibfnamefont {M.}~\bibnamefont {Wei}}, \bibinfo
		{author} {\bibfnamefont {W.}~\bibnamefont {Peng}}, \bibinfo {author}
		{\bibfnamefont {D.}~\bibnamefont {Hou}}, \bibinfo {author} {\bibfnamefont
			{Y.}~\bibnamefont {Gao}},\ and\ \bibinfo {author} {\bibfnamefont
			{Q.}~\bibnamefont {Niu}},\ }\bibfield  {title} {\bibinfo {title} {Multipolar
			anisotropy in anomalous {Hall} effect from spin-group symmetry breaking},\
	}\href {https://doi.org/10.1103/PhysRevX.15.031006} {\bibfield  {journal}
		{\bibinfo  {journal} {Phys. Rev. X}\ }\textbf {\bibinfo {volume} {15}},\
		\bibinfo {pages} {031006} (\bibinfo {year} {2025})}\BibitemShut {NoStop}%
	\bibitem [{\citenamefont {Naka}\ \emph {et~al.}(2025)\citenamefont {Naka},
		\citenamefont {Motome},\ and\ \citenamefont {Seo}}]{2025_Naka_perovskites}%
	\BibitemOpen
	\bibfield  {author} {\bibinfo {author} {\bibfnamefont {M.}~\bibnamefont
			{Naka}}, \bibinfo {author} {\bibfnamefont {Y.}~\bibnamefont {Motome}},\ and\
		\bibinfo {author} {\bibfnamefont {H.}~\bibnamefont {Seo}},\ }\bibfield
	{title} {\bibinfo {title} {Altermagnetic perovskites},\ }\href
	{https://doi.org/10.1038/s44306-024-00066-9} {\bibfield  {journal} {\bibinfo
			{journal} {npj Spintronics}\ }\textbf {\bibinfo {volume} {3}},\ \bibinfo
		{pages} {1} (\bibinfo {year} {2025})}\BibitemShut {NoStop}%
	\bibitem [{\citenamefont {Srinivasan}\ and\ \citenamefont
		{Stadler}(2022)}]{2022_rev_isolator}%
	\BibitemOpen
	\bibfield  {author} {\bibinfo {author} {\bibfnamefont {K.}~\bibnamefont
			{Srinivasan}}\ and\ \bibinfo {author} {\bibfnamefont {B.~J.~H.}\ \bibnamefont
			{Stadler}},\ }\bibfield  {title} {\bibinfo {title} {Review of integrated
			magneto-optical isolators with rare-earth iron garnets for polarization
			diverse and magnet-free isolation in silicon photonics $invited$},\ }\href
	{https://doi.org/10.1364/OME.447398} {\bibfield  {journal} {\bibinfo
			{journal} {Opt. Mater. Express}\ }\textbf {\bibinfo {volume} {12}},\ \bibinfo
		{pages} {697} (\bibinfo {year} {2022})}\BibitemShut {NoStop}%
	\bibitem [{\citenamefont {Lapointe}\ \emph {et~al.}(2025)\citenamefont
		{Lapointe}, \citenamefont {Coia}, \citenamefont {Dupont},\ and\ \citenamefont
		{Vall{\'e}e}}]{2025_Lapointe_isolator}%
	\BibitemOpen
	\bibfield  {author} {\bibinfo {author} {\bibfnamefont {J.}~\bibnamefont
			{Lapointe}}, \bibinfo {author} {\bibfnamefont {C.}~\bibnamefont {Coia}},
		\bibinfo {author} {\bibfnamefont {A.}~\bibnamefont {Dupont}},\ and\ \bibinfo
		{author} {\bibfnamefont {R.}~\bibnamefont {Vall{\'e}e}},\ }\bibfield  {title}
	{\bibinfo {title} {Passive broadband faraday isolator for hybrid integration
			to photonic circuits without lens and external magnet},\ }\href
	{https://doi.org/10.1038/s41566-024-01601-0} {\bibfield  {journal} {\bibinfo
			{journal} {Nat. Photonics}\ }\textbf {\bibinfo {volume} {19}},\ \bibinfo
		{pages} {248} (\bibinfo {year} {2025})}\BibitemShut {NoStop}%
	\bibitem [{\citenamefont {Pintus}\ \emph {et~al.}(2025)\citenamefont {Pintus},
		\citenamefont {Dumont}, \citenamefont {Shah}, \citenamefont {Murai},
		\citenamefont {Shoji}, \citenamefont {Huang}, \citenamefont {Moody},
		\citenamefont {Bowers},\ and\ \citenamefont
		{Youngblood}}]{2025_Pintus_logic}%
	\BibitemOpen
	\bibfield  {author} {\bibinfo {author} {\bibfnamefont {P.}~\bibnamefont
			{Pintus}}, \bibinfo {author} {\bibfnamefont {M.}~\bibnamefont {Dumont}},
		\bibinfo {author} {\bibfnamefont {V.}~\bibnamefont {Shah}}, \bibinfo {author}
		{\bibfnamefont {T.}~\bibnamefont {Murai}}, \bibinfo {author} {\bibfnamefont
			{Y.}~\bibnamefont {Shoji}}, \bibinfo {author} {\bibfnamefont
			{D.}~\bibnamefont {Huang}}, \bibinfo {author} {\bibfnamefont
			{G.}~\bibnamefont {Moody}}, \bibinfo {author} {\bibfnamefont {J.~E.}\
			\bibnamefont {Bowers}},\ and\ \bibinfo {author} {\bibfnamefont
			{N.}~\bibnamefont {Youngblood}},\ }\bibfield  {title} {\bibinfo {title}
		{Integrated non-reciprocal magneto-optics with ultra-high endurance for
			photonic in-memory computing},\ }\href
	{https://doi.org/10.1038/s41566-024-01549-1} {\bibfield  {journal} {\bibinfo
			{journal} {Nat. Photonics}\ }\textbf {\bibinfo {volume} {19}},\ \bibinfo
		{pages} {54} (\bibinfo {year} {2025})}\BibitemShut {NoStop}%
	\bibitem [{\citenamefont {Luo}\ \emph {et~al.}(2025)\citenamefont {Luo},
		\citenamefont {Zhou}, \citenamefont {Liang}, \citenamefont {Wang},
		\citenamefont {Zhou}, \citenamefont {Jiang}, \citenamefont {Wang},
		\citenamefont {Yao}, \citenamefont {Yang},\ and\ \citenamefont
		{Jiang}}]{2025_Luo_arxiv}%
	\BibitemOpen
	\bibfield  {author} {\bibinfo {author} {\bibfnamefont {J.}~\bibnamefont
			{Luo}}, \bibinfo {author} {\bibfnamefont {X.}~\bibnamefont {Zhou}}, \bibinfo
		{author} {\bibfnamefont {J.}~\bibnamefont {Liang}}, \bibinfo {author}
		{\bibfnamefont {L.}~\bibnamefont {Wang}}, \bibinfo {author} {\bibfnamefont
			{Q.}~\bibnamefont {Zhou}}, \bibinfo {author} {\bibfnamefont {Y.}~\bibnamefont
			{Jiang}}, \bibinfo {author} {\bibfnamefont {W.}~\bibnamefont {Wang}},
		\bibinfo {author} {\bibfnamefont {Y.}~\bibnamefont {Yao}}, \bibinfo {author}
		{\bibfnamefont {L.}~\bibnamefont {Yang}},\ and\ \bibinfo {author}
		{\bibfnamefont {W.}~\bibnamefont {Jiang}},\ }\bibfield  {title} {\bibinfo
		{title} {Symmetry-driven giant magneto-optical kerr effects in altermagnet
			hematite},\ }\href {https://arxiv.org/abs/2512.09451} {\  (\bibinfo {year}
		{2025})},\ \Eprint {https://arxiv.org/abs/2512.09451} {arXiv:2512.09451}
	\BibitemShut {NoStop}%
\end{thebibliography}
\end{document}



\title{Experimental Evidence of Néel-order-driven Magneto-optical Kerr Effect in an Altermagnetic Insulator\\ \line(1,0){60} \\ Supplementary Material}


\author{Haolin Pan}
\affiliation{International Center for Quantum Design of Functional Materials, Hefei National Research Center for Physical Sciences at the Microscale, University of Science and Technology of China, Hefei 230026, China}

\author{Rui-Chun Xiao}
\affiliation{Institute of Physical Science and Information Technology, Anhui University, Hefei 230601, China}
\affiliation{Anhui Provincial Key Laboratory of Magnetic Functional Materials and Devices, School of Materials Science and Engineering, Anhui University, Hefei 230601, China}

\author{Jiahao Han}
\affiliation{Center for Science and Innovation in Spintronics, Tohoku University, Sendai 980-8577, Japan}

\author{Hongxing Zhu}
\author{Junxue Li}
\affiliation{Department of Physics, Southern University of Science and Technology, Shenzhen 518055, China}

\author{Qian Niu}
\affiliation{CAS Key Laboratory of Strongly-Coupled Quantum Matter Physics, University of Science and Technology of China, Hefei, Anhui 230026, China}
\affiliation{Department of Physics, University of Science and Technology of China, Hefei 230026, China}

\author{Yang Gao}
\affiliation{International Center for Quantum Design of Functional Materials, Hefei National Research Center for Physical Sciences at the Microscale, University of Science and Technology of China, Hefei 230026, China}
\affiliation{CAS Key Laboratory of Strongly-Coupled Quantum Matter Physics, University of Science and Technology of China, Hefei, Anhui 230026, China}
\affiliation{Department of Physics, University of Science and Technology of China, Hefei 230026, China}
\affiliation{Anhui Center for Fundamental Sciences in Theoretical Physics, University of Science and Technology of China, Hefei 230026, China}

\author{Dazhi Hou}
\email[Correspondence author: ]{dazhi@ustc.edu.cn}
\affiliation{International Center for Quantum Design of Functional Materials, Hefei National Research Center for Physical Sciences at the Microscale, University of Science and Technology of China, Hefei 230026, China}
\affiliation{Department of Physics, University of Science and Technology of China, Hefei 230026, China}



\maketitle

\renewcommand{\thefigure}{S\arabic{figure}}  
\renewcommand{\theequation}{S\arabic{equation}}
\section{Sample Characterization}

\begin{figure}
	\centering
	\includegraphics[width=150mm]{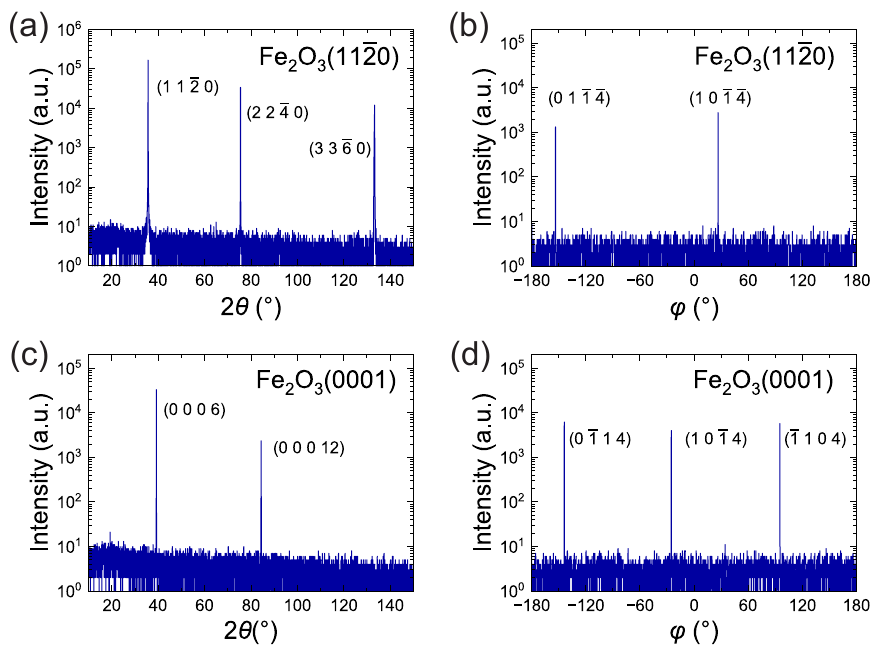}
	\caption{
		X-ray diffraction data of $\alpha$-$\rm Fe_2O_3$ $(11\bar{2}0)$ and $(0001)$ bulk samples.
		(a), (c) $2\theta$-$\omega$ scans of each sample along out-of-plane direction. 
		(b), (d) $\phi$ scans with $2\theta=33.15^\circ$. $\chi$ is $57.56^\circ$ for the $(11\bar{2}0)$ sample and $38.2^\circ$ for the $(0001)$ sample. 
	}
	\label{figS1}  
\end{figure}

X-ray diffraction data of $\rm Fe_2O_3$ $(11\bar{2}0)$ and $(0001)$ bulk samples are shown in FIG.~\ref{figS1}. The $2\theta$–$\omega$ scans in Fig.~\ref{figS1}a and \ref{figS1}c only show $(H\ H\ 2\bar{H}\ 0)$ and $(0\ 0\ 0\ 6H)$ peaks, respectively, confirming the high-quality out-of-plane monocrystallinity for each sample. The in-plane monocrystallinity is verified by scanning the $(10\bar{1}4)$ peaks as a function of the azimuthal angle $\varphi$ at appropriate tilt angles $\chi$ from the out-of-plane direction, as shown in FIG.~\ref{figS1}b and \ref{figS1}d. For the $(11\bar{2}0)$ and $(0001)$ samples, $\chi = 57.56^\circ$ and $38.2^\circ$, respectively. These results demonstrate the high quality of the single crystal samples used in this study. 

\section{Linear Response Theory for Ordinary MOKE}

We will now derive the general theory for the MOKE induced by external magnetic field, also named as ordinary MOKE. The key is to derive the following response function for the current density:
\begin{align}
	J_i=\sigma_{ijk}(\omega) E_j(\omega) B_k\,.
\end{align}
The electromagnetic field is usually induced through the magnetic vector potential. In our case, we shall introduce two vector potentials, one varies in time, i.e., $A_j^0(t)=A_j^0(\omega) e^{-i\omega t}$, and the other one varies in space, i.e., $A_\ell^\prime(\bm r)=A_\ell^\prime(\bm q) e^{i\bm q\cdot \bm r}$. The first one account for the electric field and the second one accounts for the magnetic field. The unperturbed Hamiltonian generally has the form $\hat{H}_0=\hat{H}_0(\hat{\bm p};\bm r)$. The magnetic vector potential then enters the Hamiltonian through the Peierls substitution, which shifts the momentum : $\hat{\bm p}\rightarrow \hat{\bm p}+e\bm A$ \cite{2010_Book_Condensed}. After expanding the Hamiltonian up to first order, we have~(for simple notation, we have taken $e=\hbar=1$):
\begin{align}
	\hat{H}=\hat{H}_0+\hat{\bm v}\cdot \bm A\,,
\end{align}
where $\bm A=\bm A^\prime+\bm A^0$.
We then seek the following response function
\begin{align}
	J_i=\xi_{ij\ell}(\omega,\bm q)A_j(\omega)^0 A_\ell^\prime\,.
\end{align}
By expanding the response coefficient up to first order in $\bm q$, we have
\begin{align}
	\xi_{ij\ell}(\omega, \bm q)&=\xi_{ij\ell}(\omega,0)+q_a \partial_{q_a}\xi_{ij\ell}(\omega,\bm q)|_{\bm q\rightarrow 0}+\cdots\,. 
\end{align}
Here and hereafter the repeated indices imply summation. Keeping the linear order term, we then have
\begin{align}
	J_i=\partial_{q_a}\xi_{ij\ell}(\omega,\bm q)|_{\bm q\rightarrow 0}q_a A_j(\omega)^0 A_\ell^\prime\,.
\end{align}
We then perform the antisymmetrization for $q_a$ and $A_\ell^\prime$. Since $\bm \nabla\times \bm A^\prime=i\bm q\times \bm A^\prime=\bm B$, we have
\begin{align}
	J_i=\frac{-i}{2}\epsilon_{ka\ell}\partial_{q_a}\xi_{ijk}(\omega,\bm q)|_{\bm q\rightarrow 0} A_j(\omega)^0 B_k\,.
\end{align}
Moreover, from the equation $\bm E=-\bm \nabla\phi-\partial \bm A/\partial t$, we have $\bm E=i\omega \bm A^0$.
Finally, we have
\begin{align}\label{eq_con}
	\sigma_{ijk}=\frac{-1}{2\omega}\epsilon_{ka\ell}\lim_{\bm q\rightarrow 0} \frac{\partial \xi_{ij\ell}(\omega ,\bm q)}{\partial q_a}
\end{align}

The response function $\xi_{ij\ell}$ can be derived using the density-matrix perturbation theory, which is highly successful in the derivation of optical currents. We start from the following equation of motion for the density operator $\hat{\rho}$:
\begin{align}\label{eq_sup_eom}
	\frac{d\hat{\rho}}{dt}=-i[\hat{H},\hat{\rho}]\,.
\end{align}
The Hamiltonian has two parts
\begin{align}
	\hat{H}=\hat{H}_0+\hat{H}^\prime\,,
\end{align}
where $\hat{H}_0$ is the unperturbed Hamiltonian and $\hat{H}^\prime$ is the perturbation which has the following form
\begin{align}
	\hat{H}^\prime=\sum_{i=1}^2\bm A(\omega_i,\bm q_i)\cdot \hat{\bm v}e^{-i\omega_i t}\,.
\end{align}

We expand the density operator with respect to the vector potential: $\rho=\rho_0+\delta \rho^{(1)}+\delta \rho^{(2)}+\cdots$. At the first order, we have
\begin{align}
	\frac{d\delta \rho^{(1)}}{dt}=-i[\hat{H}_0,\delta \rho^{(1)}]-i[\hat{H}^\prime,\rho_0]\,.
\end{align}
The solution reads
\begin{align}
	\delta\rho^{(1)}=-i A_j(\omega_i,\bm q_i)e^{-i\omega_i t} \int_0^\infty d\tau e^{-i\hat{H}_0\tau}[\hat{v}_j e^{i\bm q_i \cdot \bm r},\hat{\rho}_0]e^{i\hat{H}_0\tau}e^{i\omega_i\tau}\,,
\end{align}
where $\hat{\rho}_0$ is the density operator for the unperturbed system with the Hamiltonian $\hat{H}_0$.  At second order, we have
\begin{align}
	\frac{d\delta \rho^{(2)}}{dt}=-i[\hat{H}_0,\delta \rho^{(2)}]-i[\hat{H}^\prime,\rho^{(1)}]\,.
\end{align}
The solution reads
\begin{align}
	\delta\rho^{(2)}=-A_k(\omega_i,\bm q_i)A_\ell(\omega_j,\bm q_j)e^{-i(\omega_i+\omega_j)t}\int_0^\infty d\tau_1d\tau_2 e^{-i\hat{H}_0\tau_1}[\hat{v}_k e^{i\bm q_i \cdot \bm r},[\hat{v}_\ell(-\tau_2) e^{i\bm q_j\cdot \bm r},\hat{\rho}_0]] e^{i\hat{H}_0\tau_1}e^{i(\omega_i+\omega_j)\tau_1}e^{i\omega_j\tau_2}\,,
\end{align}
where $\hat{v}_k(t)=e^{i\hat{H}_0t}\hat{v}_je^{-i\hat{H}_0t}$. 
With the second order correction to the density matrix, we can then calculate the current at second order of the magnetic vector potential, i.e.,
\begin{align}
	J_i=-{\rm Tr} (\hat{\rho}^{(2)} \hat{\bm v})\,.
\end{align}

According to Eq.~\eqref{eq_con}, we then obtain the final result for the optical conductivity (here we have recovered the correct unit and the integration over crystal momentum is implied for all terms)
\begin{align}\label{eq_final}
	\sigma_{\alpha\beta\gamma}&=e^3\frac{i}{2\omega}\epsilon_{\ell k\gamma} \sum_{n_1,n_3,n_2\neq n_4}\frac{f_{n_1n_3}}{\hbar\omega+\omega_{n_1,n_3}} \frac{1}{\hbar\omega+\omega_{n_1n_2}}I_1+e^3\frac{i}{2\omega}\epsilon_{\ell k\gamma} \sum_{n_1\neq n_4,n_3,n_2}\frac{f_{n_3n_2}}{\hbar\omega+\omega_{n_3n_2}} \frac{1}{\hbar\omega+\omega_{n_1n_2}}I_2\notag\\
	&e^3 \frac{i}{2\omega}\epsilon_{\ell k\gamma} \sum_{n_1,n_3,n_2}\frac{f_{n_1n_3}}{\hbar\omega+\omega_{n_1,n_3}}\frac{(v_\ell)_{n_2n_2}}{(\hbar \omega+\omega_{n_1n_2})^2} \langle n_1|v_\alpha|n_2\rangle\langle n_2|v_k|n_3\rangle\langle n_3|v_\beta|n_1\rangle\notag\\
	&-e^3 \frac{i}{2\omega}\epsilon_{\ell k\gamma} \sum_{n_1,n_3,n_2}\frac{f_{n_3n_2}}{\hbar\omega+\omega_{n_3n_2}}\frac{(v_\ell)_{n_1n_1}}{(\hbar\omega+\omega_{n_1n_2})^2} \langle n_1|v_\alpha|n_2\rangle\langle n_2|v_k|n_3\rangle\langle n_3|v_\beta|n_1\rangle\notag\\
	&-e^3\frac{i}{2\omega}\epsilon_{\ell k\gamma} \sum_{n_1\neq n_4,n_3,n_2}\frac{f_{n_1n_3}}{\omega_{n_1n_3}} \frac{1}{\hbar \omega+\omega_{n_1n_2}}I_2-e^3\frac{i}{2\omega}\epsilon_{\ell k\gamma} \sum_{n_1,n_3,n_2\neq n_4}\frac{f_{n_3n_2}}{\omega_{n_3n_2}} \frac{1}{\hbar\omega+\omega_{n_1n_2}}I_1\notag\\
	&+ e^3\frac{i}{2\omega}\epsilon_{\ell k\gamma} \sum_{n_1\neq n_3,n_2}\left[\frac{f_{n_1n_3}}{\omega_{n_1 n_3}}\frac{(v_\ell)_{n_1n_1}}{(\hbar\omega+\omega_{n_1n_2})^2}+\frac{f_{n_1n_3}}{\omega_{n_1 n_3}^2}\frac{(v_\ell)_{n_1n_1}}{\hbar\omega+\omega_{n_1n_2}}\right] \langle n_1|v_\alpha|n_2\rangle\langle n_2|v_\beta|n_3\rangle\langle n_3|v_k|n_1\rangle\notag\\
	&- \frac{i}{2\omega}\epsilon_{\ell k\gamma} \sum_{n_1,n_3,n_2}\left[\frac{f_{n_3n_2}}{\omega_{n_3n_2}}\frac{(v_\ell)_{n_2n_2}}{(\hbar\omega+\omega_{n_1n_2})^2}+\frac{f_{n_3n_2}}{\omega_{n_3 n_2}^2}\frac{(v_\ell)_{n_2n_2}}{(\hbar\omega+\omega_{n_1n_2})^2}\right] \langle n_1|v_\alpha|n_2\rangle\langle n_2|v_k|n_3\rangle\langle n_3|v_\beta|n_1\rangle\,.
\end{align}
where
\begin{align}
	I_1&=\langle n_1|v_\alpha|n_4\rangle\langle n_4|\partial_\ell|n_2\rangle\langle n_2|v_k|n_3\rangle \langle n_3|v_\beta|n_1\rangle+\langle n_1|v_\alpha|n_2\rangle\langle \partial_\ell n_2|n_4\rangle\langle n_4|v_k|n_3\rangle \langle n_3|v_\beta|n_1\rangle\notag\\
	I_2&=\langle \partial_\ell n_1|n_4\rangle\langle n_4|v_\alpha|n_2\rangle\langle n_2|v_\beta|n_3\rangle \langle n_3|v_k|n_1\rangle+\langle n_1|v_\alpha|n_2\rangle\langle n_2|v_\beta|n_3\rangle\langle n_3|v_k|n_4\rangle \langle n_4|\partial_\ell|n_1\rangle\notag\\
	\omega_{n_1n_2}&=\varepsilon_{n_1}-\varepsilon_{n_2}\,,
\end{align}
and $\varepsilon_{n_1}$ is the band energy for the $n_1$-th band, allowing quantitative evaluation of the H-induced MOKE based on first-principles calculation.

\section{First-Principles Calculation}

The first-principles calculations based on DFT are performed using the VASP \cite{1996_Kresse_PRB, 1999_Kresse_PRB} software package. The general gradient approximation (GGA) according to the Perdew-Burke-Ernzerhof (PBE) functional is employed, and the spin-orbit coupling effects are considered. $U_{\rm eff}=4.0\ \rm eV$ is set for Fe atoms. The Bloch wave functions are iteratively transformed into maximally localized Wannier functions using the Wannier90 package \cite{2008_Mostofi_CPC}, then the effective tight binding Hamiltonian are obtained.

\begin{figure}
	\centering
	\includegraphics[width=150mm]{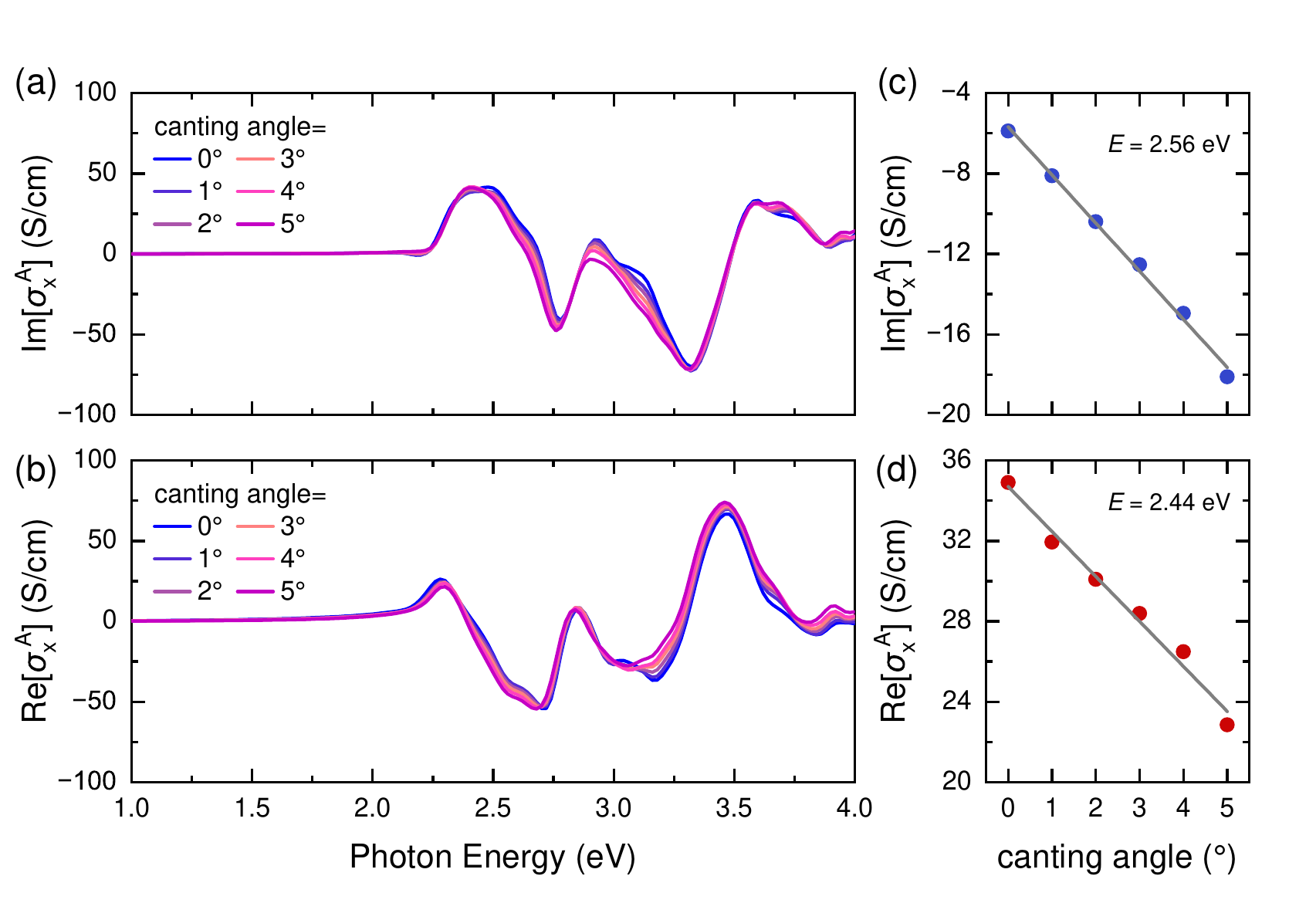}
	\caption{
		Calculation results of magnetic-order-related transverse optical conductivity $\sigma^A_x$.
		(a), (b) The real and imaginary part of $\sigma^A_x$ with different canting angles. 
		(c), (d) Demonstrations of linear fitting at specific photon energy. The slopes correspond to the MO coefficients of net magnetization.
	}
	\label{figS2}  
\end{figure}

According to Kubo-Greenwood theory \cite{1996_Ebert_RPP,2004_Yao,2016_Xiao_PRL}, we compute the magnetic-order-related transverse optical conductivity, containing the contributions of N\'eel order and net magnetization, as follows: 
\begin{align}\label{eq_KG}
\sigma^A_k(\omega)= \frac{\epsilon_{ijk}}{2}\hbar e^2 \int  \frac{d^3\bf{k}}{(2\pi)^3}\sum_{n\neq m} \frac{f_{m\bf{k}}-f_{n\bf{k}}}{\varepsilon_{m\bf{k}}-\varepsilon_{n\bf{k}}} \frac{{\rm Im} \langle\psi_{n\bf{k}}|\nu_i|\psi_{m\bf{k}}\rangle\langle\psi_{m\bf{k}}|\nu_j|\psi_{n\bf{k}}\rangle}{\varepsilon_{m\bf{k}}-\varepsilon_{n\bf{k}}-(\hbar\omega+i\eta)},
\end{align}
where $\omega$ is the frequency of light, $\epsilon_{ijk}$ is the Levi-Civita symbol. In specific, we calculate $\sigma^A_x$ in Eq.~\ref{eq_KG} considering $\vec N \parallel y$. To separate the two contributions, the spin arrangements with different canting angles, from 0 to 5 degrees, are considered in calculations. The results are shown in FIG.~\ref{figS2}a and \ref{figS2}b. The contribution of N\'eel order, $\alpha N$, equals to $\sigma^A_x$ when canting angle is zero. The MO coefficients of net magnetization, $\beta_{xx}$, is obtained by linearly fitting the slope of $\sigma^A_x$ against canting angle at each photon energy, as demonstrated in FIG.~\ref{figS2}c and \ref{figS2}d. $\beta_{xx}$ follows $\beta_{xx}=\rm{slope}/(2M_{sub}*\pi/180)$, where $\rm M_{sub}=7.75\times 10^{5}\ A/m$ is the sublattice magnetization of hematite. For comparison, the contribution of spontaneous magnetization, $\beta_{xx}M_S$, is shown in the main text, where $M_S=2.13\times10^3\rm\ A/m$.

The coefficient $\gamma_{xx}$ induced by external magnetic field is calculated based on Eq.~\ref{eq_final}, with $\gamma_{xx}=\sigma_{yzx}$. For comparison, the contribution of external magnetic field, $\gamma_{xx}H_0$, is shown in the main text, where $H_0=1\rm\ T$.

At this point, it is worth discussing the relationship between the MOKE signal of hematite and SOC. Although altermagnets can exhibit spin-band splitting without SOC, the MOKE induced by N\'eel order still requires a nonzero SOC, which has been thoroughly discussed in a recent theoretical article \cite{2026_Xiao}. In the absence of SOC, the spin and orbit of electrons are independent, and spin-band splitting does not influence the electron's motion. Only when the two interact with each other will altermagnets exhibit AHE and MOKE due to spin splitting.

\section{Details of MOKE Measurement and Data Analysis}

\begin{figure}
	\centering
	\includegraphics[width=86mm]{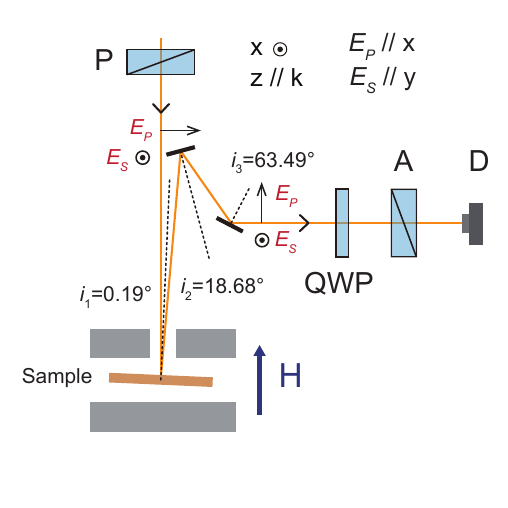}
	\caption{
		The optical path of MOKE measurement. P --- Polarizer. QWP --- Quarter Wave Plate. A --- Analyzer. D --- Detector.
	}
	\label{figS3}  
\end{figure}
The optical path of MOKE measurement is shown in FIG.~\ref{figS3}. The $xyz$ coordinate system in FIG.~\ref{figS3} is defined on the optical path, while $x'y'z'$ represents the crystal coordinate system. 
The light source is a supercontinuum laser equipped with an acousto-optic tunable filter. After being polarized to $E_S$ by a Glan-Taylor prism, the light is reflected sequentially by the sample and two Ag mirrors, with incident angles $i_1 = 0.19^\circ$, $i_2 = 18.68^\circ$, and $i_3 = 63.49^\circ$. 
For the $(11\bar{2}0)$ sample, the polarization of the incident light is parallel to the $[0001]$ direction, while for the $(0001)$ sample, it is parallel to the $[11\bar{2}0]$ direction.

A broadband quarter-wave plate (QWP) is used for ellipticity measurement, but it is removed from the optical path for rotation measurements. The fast axis of the QWP is aligned with the $y$-axis. The optical axis of the analyzer is positioned at an angle $\delta$ from $y+$ to $x+$. The photoelectric detector measures the intensity of the light, denoted as $I$.

The rotation angle ($\theta$) and ellipticity ($\varepsilon$) of the light after three reflections are given by \cite{2000_Qiu_MOreview}: 
$$\theta=\delta\cdot \frac{I}{2I_0},\ \ \ \varepsilon=-\delta\cdot \frac{I}{2I_0\sin\Delta}+\frac{\theta}{\tan\Delta},$$
where $I_0$ is the average intensity measured during the H-sweeping procedure and $\Delta$ is the retardation of the QWP as specified by the manufacturer.
Using the Jones formalism, the detected electric field vector is described as:
\begin{equation}
	E\left(
	\begin{array}{c}
		1 \\ 
		-\theta+i\varepsilon
	\end{array}
	\right)
	=\boldsymbol{J}_3\boldsymbol{J}_2
	\left(
	\begin{array}{cc}
		r_{ss} & r_{sp}  \\ 
		r_{ps} & r_{pp}
	\end{array}
	\right)
	\left(
	\begin{array}{c}
		E_S \\ 
		0
	\end{array}
	\right),
\end{equation}
where $\boldsymbol{J}_k$ is the Mueller matrix for the $k$-th reflection with incident angle $i_k$, $k = 2, 3$, which is given by:
$$\boldsymbol{J}_k=\left(
\begin{array}{cc}
	r^{Ag}_{ss}(i_k) & 0  \\ 
	0 & r^{Ag}_{pp}(i_k)
\end{array}
\right)$$ 
The reflection coefficients $r^{Ag}{ss}$ and $r^{Ag}{pp}$ for the Ag mirror are obtained using the Fresnel equations. Thus, the ratio $r_{ps}/r_{ss}$ for the sample is determined as:
\begin{equation}
	\frac{r_{ps}}{r_{ss}}=\frac{r^{Ag}_{ss}(i_3)r^{Ag}_{ss}(i_2)}{r^{Ag}_{pp}(i_3)r^{Ag}_{pp}(i_2)}\cdot(-\theta+i\varepsilon).
\end{equation}
As an example, consider the measurement for $\rm Fe_2O_3(11\bar{2}0)$. Using the MOKE formula for polar geometry in the thick-film limit \cite{1998_You_formula}, the ratio $r_{ps}/r_{ss}$ is given by:
\begin{equation}
	\frac{r_{ps}}{r_{ss}}=\frac{\tilde{n}}{\tilde{n}^2-1}\cdot\frac{\epsilon_{y'z'}}{\epsilon_{z'z'}},
\end{equation}
where $\tilde{n} = n - ik$ is the complex refractive index of hematite, and $\epsilon_{z'z'} = n^2 - k^2 + 2nki$. The transverse optical conductivity $\sigma^A_{x'}$ is related to the dielectric tensor by $\epsilon_{y'z'} = i \sigma^A_{x'} / \omega$, where $\omega$ is the light frequency. The complex refractive indices of Ag and hematite are taken from previously reported values \cite{1985_Querry_N, 1972_Johnson_N}, for which we use extraordinary-light data for $(11\bar{2}0)$ samples and ordinary-light data for $(0001)$ sample. Using this approach, we obtain the absolute value of the transverse optical conductivity, as presented in the main text.


%